\documentclass[a4paper, 12pt]{article}
\usepackage{graphicx}
\usepackage{cite}

\begin{document}

\def \d {{\rm d}}
\def \im {{\rm i}}
\def \boldk {\mbox{\boldmath$k$}}
\def \boldl {\mbox{\boldmath$l$}}
\def \boldm {\mbox{\boldmath$m$}}
\def \boldu {\mbox{\boldmath$u$}}
\def \boldn {\mbox{\boldmath$n$}}
\def \boldv {\mbox{\boldmath$v$}}
\def \bolda {\mbox{\boldmath$a$}}

\title{Photon rockets moving arbitrarily in any dimension}

\author{Ji\v{r}\'{\i} Podolsk\'y \thanks{podolsky@mbox.troja.mff.cuni.cz}
\\ \\Institute of Theoretical Physics, Faculty of Mathematics and Physics,\\
 Charles University in Prague, \\
 V Hole\v{s}ovi\v{c}k\'{a}ch 2, 180 00 Prague 8,  Czech Republic}

\date{\today}

\maketitle

\abstract{A family of explicit exact solutions of Einstein's equations in four and higher dimensions is studied which describes the gravitational field of an object accelerating due to an anisotropic emission of photons. It is possible to prescribe an arbitrary motion, so that the acceleration of such photon rocket need not be uniform --- both its magnitude and direction may vary with time. Except at location of the rocket the spacetimes have no curvature singularities, and topological defects like cosmic strings are also absent. Any value of a cosmological constant is allowed. We investigate some particular examples of motion, namely a straight flight and a circular trajectory, and we derive the corresponding radiation patterns and the mass loss of the rockets. We also demonstrate the absence of ``gravitational aberration'' in such spacetimes. This interesting member of the higher-dimensional Robinson--Trautman class of pure radiation spacetimes of algebraic type~D generalises the class of Kinnersley's solutions that has long been known in four-dimensional general relativity.
}

\vspace{.2cm}


\section{Introduction}
\label{intro}

In 1969, Kinnersley~\cite{Kin69} introduced and investigated a class of exact spacetimes that describe a localised object which accelerates due to the back reaction of the emitted null radiation. In the axially symmetric case the object moves along a straight line corresponding to the axis, but more general solutions also exist in which the motion is completely arbitrary. They may serve as interesting  self-consistent exact models for the motion a photon rocket that is propelled by a specific anisotropic emission of photons. Such solutions have attracted considerable attention \cite{Bon94,Dam95,Bon96,DaiMorGle96,CorMic96,vonGonKram98,Cor00,Car00,Pod08}, in particular due to some unusual properties of the associated gravitational radiation.

The Kinnersley solutions are of algebraic type~D and belong to a large family of Robinson--Trautman spacetimes \cite{RobTra60,RobTra62,Stephanibook,GriPod09} which is defined by the property that it admits a geodesic, shear-free, twist-free but expanding null vector field. Interestingly, within the class of Robinson--Trautman spacetimes there are other distinct type~D solutions which also describe accelerated sources, namely the famous \hbox{C-metric}. This is a specific vacuum solution which represents a pair of black holes, uniformly accelerating under the influence of cosmic strings or struts along the axis of symmetry~\cite{KinWal70} (for a recent review see \cite{GriPod09}). In contrast to the \hbox{C-metric}, the spacetimes describing Kinnersley's photon rockets are non-vacuum (the rockets emit pure radiation) and, apart from the location of the rocket, they are regular everywhere.

It is desirable to find and study higher-dimensional extensions of such classes of exact spacetimes which represent accelerated objects. There have been many attempts to generalise the C-metric to higher dimensions, but it has not (yet) been found. Interestingly enough, an exact class of pure radiation spacetimes which includes those of the Kinnersley rockets in an arbitrary dimension has been recently discovered within the Kerr--Schild family~\cite{GurSar02,GurSar04} and independently in the Robinson--Trautman family~\cite{PodOrt06}.

It is the purpose of this contribution to present and analyse such spacetimes. In section~\ref{Krocketsection} we briefly summarise the Kinnersley photon rockets in four dimensions. Subsequently, in section~\ref{trajMink} we introduce very useful Newman--Unti coordinates adapted to an arbitrarily moving test particle in a flat space of dimension $D$. In section~\ref{propsection} we present a generalisation of the Kinnersley solution to higher dimensions, discussing the Kerr--Schild and the Robinson--Trautman forms and a possible cosmological constant. Sections~\ref{singlesection} and~\ref{examplsection} concentrate on particular situations in which the rockets accelerate along a single spatial direction or move along a circular trajectory. In both these cases we derive and plot the corresponding radiation pattern and the mass loss formula. In the final section~\ref{abersection} we give the Christoffel symbols necessary for study of geodesics in these spacetimes and we demonstrate the absence of ``gravitational aberration''.

\section{Kinnersley photon rockets in ${D=4}$}
\label{Krocketsection}

Kinnersley's solution~\cite{Kin69} belongs to the family of  Robinson--Trautman space-times with an aligned pure radiation. In four dimensions, the metric can thus be written in the standard form \cite{RobTra60,RobTra62,Stephanibook,GriPod09}
 \begin{equation}
 \d s^2 =2\,\frac{r^2}{P^2}\,\d\zeta\,\d\bar\zeta -2\,\d u\,\d r  -2\,H\,\d u^2,
 \label{RTmetric}
 \end{equation}
 where
 \begin{equation}
 2H = K -2r(\,\log P)_{,u} -\frac{2m(u)}{r} -\frac{\Lambda}{3}r^2.
 \label{RTHfunction}
 \end{equation}
Here $\Lambda$ is the  cosmological constant while $P$ is an arbitrary function ${P(u,\zeta,\bar\zeta)}$. The function ${\,K(u,\zeta,\bar\zeta)\equiv\Delta\,\log P}$, where ${\Delta\equiv 2P^2\partial_{\zeta}\partial_{\bar\zeta}\,}$, determines the Gaussian curvature of the 2-surfaces spanned by the (complex) spatial coordinate~$\zeta$, on which ${r=1}$ and $u$ is any constant.

The family of Kinnersley rockets is obtained when
 \begin{equation}
  P=A(u)+B(u)\, \zeta+\bar B(u)\,\bar\zeta+C(u)\,\zeta\bar\zeta\,,
 \label{generalKinnersley}
 \end{equation}
in which ${A,B,C}$ are functions of $u$ ($B$ may be complex), so that
 \begin{equation}
 K(u)=2(AC-B\bar B)\,.
 \label{KgeneralKinnersley}
 \end{equation}
For ${K>0}$, these solutions represent the gravitational field of an arbitrarily moving object located at ${r=0}$ whose velocity is encoded in the functions ${A,B,C}$ (for the explicit relations see equation (\ref{functionP2stereo}) below; more details are given in section~\ref{trajMink}). Since the Gaussian curvature $K(u)$ is independent of the transverse spatial coordinates ${\zeta,\bar\zeta}$, there are no poles over the compact surfaces ${u=}$~const. (at any $r$) which means that for ${r\not=0}$ these spacetimes are everywhere regular, free of curvature singularities and cosmic strings or struts. Indeed, using the natural null tetrad ${\boldk=\partial_r}$, ${\boldl=\partial_u-H\partial_r}$, ${\boldm=(P/r)\,\partial_{\bar\zeta}}$ the only component of the curvature tensor for the metric (\ref{RTmetric}) with (\ref{generalKinnersley}) is
\begin{equation}
 \Psi_2=-\frac{m(u)}{r^3}\,.
 \label{RTPsis}
\end{equation}
The spacetimes are thus of algebraic type~D, and there is a curvature singularity only at ${r=0}$ where ${\Psi_2}$ diverges. Conformal infinity is located at ${r=\infty}$ where the space-times asymptotically become conformally flat and also vacuum, i.e.,  Minkowski or (anti-)de~Sitter, according to the sign of $\Lambda$.

The spacetimes contain an aligned pure radiation field (that is flow of matter of zero rest-mass, emitted from the source located at ${r=0}$) with an energy-momentum tensor of the form ${T_{\mu\nu}=\rho\, k_\mu k_\nu\,}$. The radiation density is
 \begin{equation}
 \rho=\frac{n^2(u,\zeta,\bar\zeta)}{r^2}\,,
 \label{pureradrho}
 \end{equation}
 where the function $n^2$ is determined by Einstein's equation as
 \begin{equation}
4\pi\, n^2 =  -m_{,u} + 3m(\,\log P)_{,u}\,.
 \label{RTequationpurerad}
 \end{equation}

In the particular case when ${B=0}$ and ${K=+1}$, the Kinnersley photon rocket moves along a single axis. Performing a suitable transformation, the metric (\ref{RTmetric})--(\ref{generalKinnersley}) then can be put into the form
 \begin{eqnarray}
 &&\hskip-0.98pc\d s^2 =-\bigg(1-{2\,m\over r} -{\Lambda\over3}r^2-2\alpha\, r\cos\vartheta -\alpha^2\,r^2\sin^2\vartheta\bigg)\d u^2  \nonumber\\
 &&\hskip4pc  -\,2\,\d u\,\d r+\>2\alpha\,r^2\sin\vartheta\,\d u\,\d\vartheta   +\,r^2(\d\vartheta^2+\sin^2\vartheta\,\d\phi^2)\,, \label{Kinnersrocketmetric}
  \end{eqnarray}
in which $\alpha(u)$ is an arbitrary ``acceleration function'' of the coordinate~$u$, see \cite{Kin69,Bon94,Pod08} for more details.
This represents a singular source located at ${r=0}$, of decreasing mass determined by $m(u)$, which emits pure radiation and accelerates in Minkowski, de~Sitter or anti-de~Sitter space along a straight line (which is the axis of symmetry) due to the corresponding net back reaction. It thus serves as a simple exact model of a rocket that is propelled by the anisotropic emission of photons whose radiation field profile is determined by the field equation (\ref{RTequationpurerad}) as ${n^2(u,\vartheta) = \frac{1}{4\pi}[-m_{,u}(u)+3\,\alpha(u)\, m(u)\cos\vartheta]}$.

\section{Arbitrarily accelerated coordinates in flat space}
\label{trajMink}

Before presenting a generalisation of the Kinnersley photon rockets solution to higher dimensions (in the following section~\ref{propsection}), it will be important to introduce and summarise a particularly useful coordinate system adapted to an arbitrarily moving test particle in a flat space of dimension $D$. For ${D=4}$ this was described already by Newman and Unti in their classic work \cite{NewUnt63}, and subsequently elsewhere.

\begin{figure}[ht]
   \begin{center}
   \includegraphics[scale=0.70]{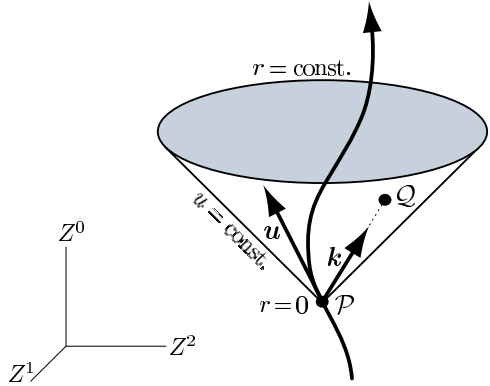}
   \end{center}
   \caption{\label{fig1}%
   Construction of the coordinate system that is naturally adapted to an arbitrarily moving test particle in $D$-dimensional flat space (spatial coordinates ${Z^3,\ldots,Z^{D-1}}$ are suppressed). At any event ${\cal P}$ of the particle's worldline with velocity $\boldu$, the future null cone is labeled as $u$ and it is assigned the value of the corresponding proper time. The coordinate $r$ is an affine parameter along null geodesics which connect ${\cal P}$ with any event ${\cal Q}$ on the cone. These are generated by properly normalised null vectors $\boldk$. }
   \end{figure}

Let us consider Minkowski space of dimension $D$ with the standard metric
 \begin{equation}
 \d s^2=\eta_{\alpha\beta}\,\d Z^\alpha \d Z^\beta=-{(\d Z^0)}^2+{(\d Z^1)}^2+\cdots +{(\d Z^{D-1})}^2 ,
 \label{MinkowskiD}
 \end{equation}
and \emph{any timelike worldline}
 \begin{equation}
 z^\alpha(u)\,.
 \label{worldline}
 \end{equation}
This describes motion of a test particle in the flat space (\ref{MinkowskiD}), and the parameter~$u$ is assumed to be its proper time. The corresponding velocity is
 \begin{equation}
 \boldu=\dot z^\alpha\,\partial_\alpha\,,
 \quad\hbox{where}\quad \dot z^\alpha(u)\equiv \frac{\d z^\alpha}{\d u\,}
 \quad\hbox{and}\quad \partial_\alpha\equiv\frac{\partial}{\partial Z^\alpha}\,.
 \label{velocity}
 \end{equation}
At any event ${{\cal P}\equiv\{z^0(u), \ldots, z^{D-1}(u)\}}$ on the trajectory (\ref{worldline}) we now take the \emph{future null cone} and we label this hypersurface as ${u=\,}$const. which is exactly the value of the proper time of the particle when it passes through the vertex point ${\cal P}$, as shown in figure~\ref{fig1}.

\noindent
Each of such null cones is generated by a family of null vectors ${\boldk=k^\alpha\,\partial_\alpha}$ along the particle's trajectory. It is convenient to prescribe the normalisation condition ${\boldk\cdot\boldu=-1}$, in addition to those already mentioned, ${\boldu\cdot\boldu=-1}$ and ${\boldk\cdot\boldk=0}$, i.e.:
 \begin{eqnarray}
 \eta_{\alpha\beta}\,k^\alpha \dot z^\beta &=& -1\,, \label{kunormaliz}\\
 \eta_{\alpha\beta}\,\dot z^\alpha \dot z^\beta &=& -1\,, \label{uunormaliz}\\
 \eta_{\alpha\beta}\,k^\alpha k^\beta &=& 0\,. \label{kknormaliz}
 \end{eqnarray}
The background coordinates of any event ${{\cal Q}\equiv\{Z^0, \ldots, Z^{D-1}\}}$ on the null cone ${u=\,}$const. can obviously be expressed as ${Z^\alpha=z^\alpha(u)}+ r\, k^\alpha$, where $k^\alpha$ represents the generator of the null geodesics connecting ${\cal Q}$ with ${\cal P}$, and $r$ is a new coordinate. In fact, $r$ is the corresponding \emph{affine parameter} since ${\partial_r=Z^\alpha_{\,\,,r}\,\partial_\alpha=k^\alpha\partial_\alpha=\boldk}$, and using the relation (\ref{kunormaliz}) this can be explicitly expressed as ${r=-\eta_{\alpha\beta}(Z^\alpha-z^\alpha(u))\,\dot z^\beta}$. Now it only remains to parameterise the null vectors $\boldk$. In the natural frame of the background coordinates of (\ref{MinkowskiD}), without loss of generality all these vectors can be conveniently represented as
 \begin{equation}
 \boldk=\frac{1}{p(u,\boldn)}\,\Big(\,\partial_0+\sum_{i=1}^{D-1}n^i\,\partial_i\,\Big),
 \label{represk}
 \end{equation}
i.e., ${k^0=p^{-1}, k^i=p^{-1}n^i}$. Here $p(u,\boldn)$ is some function, and the parameters $n^i$ can be understood as components of a \emph{spatial unit vector} ${\boldn=n^i\partial_i}$. Indeed,
 \begin{equation}
 \sum_{i,j=1}^{D-1}\delta_{ij}\,n^i n^j=1
 \label{normn}
 \end{equation}
guarantees that ${\boldk\cdot\boldk=0}$.

The natural geometrical construction described above thus introduces new convenient set of $D$ independent coordinates in the flat spacetime, namely $u$, $r$ and $n^i$. Recall that there are ${D-1}$ spatial parameters $n^i$, but since they are constrained by the normalisation condition (\ref{normn}), only ${D-2}$ of them are independent. These new coordinates are adapted to an arbitrarily moving (accelerating) test particle which is always located at ${r=0}$, the vertex of all future-oriented null cones. They are labelled by the coordinate $u$ which has the meaning of a retarded time because it coincides with the proper time of the moving particle.

Now it is straightforward to express the Minkowski space (\ref{MinkowskiD}) in terms of these new coordinates ${\{u, r, n^i\}}$ using the relation
 \begin{equation}
 Z^\alpha=z^\alpha(u)+ r\, k^\alpha(u,\boldn)\,.
 \label{transformation}
 \end{equation}
Employing relations (\ref{kunormaliz}), (\ref{uunormaliz}), (\ref{kknormaliz}) and also the identities
 \begin{eqnarray}
 2\,\eta_{\alpha\beta}\,k^\alpha \d k^\beta &=& \d ( \eta_{\alpha\beta}\,k^\alpha k^\beta) =0\,, \label{id1}\\
 \eta_{\alpha\beta}\,\dot z^\alpha \d k^\beta &=& (\,\log p)_{,u}\,\d u\,, \label{id2}\\
 \eta_{\alpha\beta}\,\d k^\alpha \d k^\beta &=& \frac{1}{p^2}\sum_{i,j=1}^{D-1}\delta_{ij}\, \d n^i \d n^j \,, \label{id3}
 \end{eqnarray}
we arrive at the following metric form
 \begin{equation}
 \d s^2 =\frac{r^2}{p^2}\sum_{i,j=1}^{D-1}\delta_{ij}\, \d n^i \d n^j -2\,\d u\,\d r  - \Big(1-2r(\,\log p)_{,u}\Big)\,\d u^2,
 \label{RTMinkowskiD}
 \end{equation}
with the constraint (\ref{normn}). The function $p(u,\boldn)$ in (\ref{RTMinkowskiD}) is not arbitrary since the condition ${\boldk\cdot\boldu=-1}$, i.e. (\ref{kunormaliz}), must be satisfied. It immediately follows from (\ref{kunormaliz}) and (\ref{represk}) that
 \begin{equation}
 p(u,\boldn)=\dot z^0(u)-\sum_{i=1}^{D-1}n^i\dot z^i(u) \,,
 \label{functionP}
 \end{equation}
where ${\dot z^\alpha(u)}$ are components of the velocity (\ref{velocity}) of an arbitrarily moving particle whose worldline is ${z^\alpha(u)}$.

\subsection{Flat metric in accelerated spherical-like coordinates}
\label{trajMinkspher}
The most natural explicit parametrisation of the components $n^i$ of the unit vector $\boldn$, which satisfies (\ref{normn}), is given by ${D-2}$ spherical angles. In ${D=4}$ the standard choice
 \begin{equation}
 n^1=\cos\theta\,,\quad n^2=\sin\theta\cos\phi\,,\quad n^3=\sin\theta\sin\phi\,,
 \label{nispher}
 \end{equation}
brings the metric (\ref{RTMinkowskiD}) to the form
\begin{equation}
 \d s^2 ={r^2\over p^2}\,(\d \theta^2+\sin^2\,\theta \,\d \phi^2) -2\,\d u\,\d r  - \Big(1-2r(\,\log p)_{,u}\Big)\,\d u^2,
 \label{RTMinkowski2spher}
 \end{equation}
where, due to (\ref{functionP}),
\begin{equation}
 p(u,\theta,\phi)=\dot z^0-\dot z^1\cos\theta-\dot z^2\sin\theta\cos\phi-\dot z^3\sin\theta\sin\phi  \,.
 \label{functionP2spher}
 \end{equation}
This can further be rewritten in terms of the complex stereographic coordinate $\zeta\,$,
 \begin{equation}
 \zeta=\sqrt{2}\,\tan\frac{\theta}{2}\,e^{\im\phi}\,,
 \label{stereo}
 \end{equation}
as
\begin{equation}
 \d s^2 =2\,\frac{r^2}{P^2}\,\d\zeta\,\d\bar\zeta  - 2\,\d u\,\d r  - \Big(1-2r(\,\log P)_{,u}\Big)\,\d u^2,
 \label{RTMinkowski2stereo}
 \end{equation}
where the function ${\,P\equiv(1+\frac{1}{2}\zeta\bar\zeta)\,p\,}$ is given by
\begin{equation}
\hspace{-15mm}
{\textstyle P(u,\zeta,\bar\zeta) = (\dot z^0-\dot z^1)-\frac{1}{\sqrt 2}(\dot z^2-\im\,\dot z^3)\zeta -\frac{1}{\sqrt 2}(\dot z^2+\im\,\dot z^3)\bar\zeta + \frac{1}{2}(\dot z^0+\dot z^1)\zeta\bar\zeta}\,.
 \label{functionP2stereo}
 \end{equation}
This Robinson--Trautman form (\ref{RTmetric}), (\ref{RTHfunction}) of the flat metric, which was presented e.g. in \cite{NewUnt63,Fro76,Cor00}, gives an explicit physical meaning to the mathematical coefficients ${A(u), B(u), C(u)}$ in the function (\ref{generalKinnersley}). Notice also that ${K=1}$ due to (\ref{uunormaliz}), as required by equation (\ref{KgeneralKinnersley}).

Spherical parametrisation of the components $n^i$ of a unit vector $\boldn$ by ${D-2}$ angles~$\theta_i$ can be given in any dimension~$D$. Let us consider
 \begin{equation}
 n^i=\cos\theta_i\,\prod_{j=1}^{i-1} \sin\theta_j\,,
 \label{nispherD}
 \end{equation}
and define
 \begin{equation}
 \theta_{D-1}\equiv0\,, \qquad  \prod_{j=1}^{0} \sin\theta_j \equiv 1\,.
 \label{nispherDdef}
 \end{equation}
Relation (\ref{normn}) is identically satisfied and the spatial part of the metric (\ref{RTMinkowskiD}) becomes
 \begin{eqnarray}
 \hspace{-13mm}
 \sum_{i,j=1}^{D-1}\delta_{ij}\, \d n^i \d n^j &=& \sum_{i=1}^{D-2} \Big( \prod_{j=1}^{i-1} \sin^2\theta_j \Big) \d\theta_i^2  \label{RTMinkowskiDspat}\\
 && \hspace{-15mm} = \d\theta_1^2 + \sin^2\theta_1\,\d\theta_2^2 + \sin^2\theta_1\sin^2\theta_2\,\d\theta_3^2 + \cdots + \sin^2\theta_1\ldots\sin^2\theta_{D-3}\,\d\theta_{D-2}^2\,, \nonumber
 \end{eqnarray}
where ${\theta_1, \theta_2,\ldots,\theta_{D-3}\in[0,\pi]}$, ${\theta_{D-2}\in[0,2\pi)}$. Clearly, it has the geometry of a sphere $S^{D-2}$.
Flat space in arbitrarily accelerated coordinates thus takes the metric form
\begin{equation}
\hspace{-10mm}
 \d s^2 ={r^2\over p^2}\sum_{i=1}^{D-2} \Big( \prod_{j=1}^{i-1}  \,\sin^2\theta_j \Big) \d\theta_i^2 -2\,\d u\,\d r  - \Big(1-2r(\,\log p)_{,u}\Big)\,\d u^2,
 \label{RTMinkowskiDspher}
 \end{equation}
in which \begin{equation}
 p(u,\theta_i)=\dot z^0(u)-\sum_{i=1}^{D-1}\dot z^i(u)\,\cos\theta_i \prod_{j=1}^{i-1} \sin\theta_j \,.
 \label{functionPDspher}
 \end{equation}
Recall again that the functions ${\dot z^0(u), \dot z^1(u), \ldots, \dot z^{D-1}(u)}$ are components of the velocity (\ref{velocity}) of the test particle located at ${r=0}$ which moves along an \emph{arbitrary} worldline (\ref{worldline}), and $u$ is its proper time.

\subsection{Flat metric in accelerated Cartesian-like coordinates}
\label{trajMinkscart}
Another important parametrisation of the general metric (\ref{RTMinkowskiD}) is obtained in terms of Cartesian-type coordinates. These are introduced by
 \begin{eqnarray}
  n^i    &=& \frac{x^i}{1+\frac{1}{4}\delta_{kl}\,x^kx^l} \,,\qquad\quad i,j,k,l=1,2,\ldots, D-2\,, \nonumber\\
  n^{D-1}&=& \frac{1-\frac{1}{4}\delta_{ij}\,x^ix^j}{1+\frac{1}{4}\delta_{kl}\,x^kx^l} \,, \label{nicartD}
 \end{eqnarray}
 (with the summation convention over ${i,j,k,l}$)
 in which the spatial part of the metric becomes explicitly conformally flat. Indeed, ${\sum_{i,j=1}^{D-1}\delta_{ij}n^i n^j=1}$ and
 \begin{equation}
 \sum_{i,j=1}^{D-1}\delta_{ij}\, \d n^i \d n^j = \frac{\delta_{ij}\, \d x^i \d x^j}{(1+\frac{1}{4}\delta_{kl}\,x^kx^l)^2} \,. \label{RTMinkowskiDspatcart}
 \end{equation}
The metric (\ref{RTMinkowskiD}) thus becomes
\begin{equation}
 \d s^2 =\frac{r^2}{P^2}\,\delta_{ij}\, \d x^i \d x^j -2\,\d u\,\d r  - \Big(1-2r(\,\log P)_{,u}\Big)\,\d u^2,
 \label{RTMinkowski2cart}
 \end{equation}
in which the function
\begin{equation}
 P\equiv(1+{\textstyle\frac{1}{4}}\delta_{kl}\,x^kx^l)\,p
 \label{relationPp}
 \end{equation}
with $p$ given by (\ref{functionP}) takes the form
\begin{equation}
{\textstyle P = (\dot z^0-\dot z^{D-1})-(\delta_{ij}\,\dot z^j)\,x^i + \frac{1}{4}(\dot z^0+\dot z^{D-1})\,\delta_{ij}\, x^i x^j}\,.
 \label{functionP2cart}
 \end{equation}
This is exactly the canonical form of the $D$-dimensional Robinson--Trautman metric, as given by equations (37), (38) of \cite{PodOrt06} in the flat-space case when ${\Lambda=0}$ and ${\mu=0}$. In fact, it gives an explicit physical meaning to general coefficients in the function (42) therein,
 \begin{equation}
P=A(u)+ B_i(u)\,x^i + C(u)\, \delta_{ij}\, x^i x^j\,,
 \label{RTMinkowski2cartP}
 \end{equation}
namely
 \begin{eqnarray}
 A (u)  &\equiv& \dot z^0(u)-\dot z^{D-1}(u)\,, \nonumber\\
 B_i(u) &\equiv& -\dot z^i(u)\,, \label{RTMinkowskiDabc}\\
 C (u)  &\equiv& {\textstyle\frac{1}{4}}[\,\dot z^0(u)+\dot z^{D-1}(u)\,]\,. \nonumber
 \end{eqnarray}
Consequently, the Ricci scalar ${{\cal R}}$ corresponding to the spatial metric ${h_{ij}\equiv P^{-2}\delta_{ij}}$ is
 \begin{equation}
 \frac{{\cal R}(u)}{(D-2)(D-3)}=4AC-\sum_{i=1}^{D-2} B_i^2=-\eta_{\alpha\beta}\,\dot z^\alpha \dot z^\beta=1\,.
 \label{RTMinkowski2cartR}
 \end{equation}
The transverse $(D-2)$-dimensional space, covered by the coordinates $x^i$, thus has a constant positive curvature, i.e., it is a sphere $S^{D-2}$ of radius $r$, as in (\ref{RTMinkowskiDspher}).

In particular, when the test particle located at the origin ${r=0}$ of the coordinates of  (\ref{RTMinkowski2cart}) is \emph{at rest} with respect to the Minkowski background (\ref{MinkowskiD}) then $\dot z^0(u)=1$, ${\dot z^i(u)=0=\dot z^{D-1}(u)}$. The metric function (\ref{RTMinkowski2cartP}) thus reduces to ${P=1+ \frac{1}{4}\, \delta_{ij}\, x^i x^j}$, which gives exactly the well-known form of the Schwarzschild--Tangherlini black hole spacetime in the limit when its mass parameter vanishes.

The function (\ref{functionP2stereo}) obviously equals (\ref{functionP2cart}) for ${D=4}$ if we introduce the complex coordinate ${\zeta=\frac{1}{\sqrt2}(x^1+\im\,x^2)}$ and relabel ${z^1 \to z^3 \to z^2 \to z^1}$.

\section{Photon rockets in higher dimensions}
\label{propsection}

We will now present the class of non-flat solutions which generalise the Kinnersley spacetimes from ${D=4}$ to ${D>4}$. These contain null radiation (representing the emission of photons) and a possible cosmological constant~$\Lambda$. Such solutions describe an arbitrarily moving photon rocket in any dimension. To this end, the coordinate systems naturally adapted to general worldlines of test particles in flat space, as described in previous section~\ref{trajMink}, will be employed for the background.

\subsection{The Kerr--Schild form}
\label{Kerr_Schild}
In fact, these exact solutions of Einstein's field equations in higher dimensions have already been found by several authors. In particular,
G\"urses and Sar\i o\u{g}lu in~\cite{GurSar02,GurSar04} presented an accelerated metric which has the Kerr--Schild form
 \begin{equation}
 \d s^2=\eta_{\alpha\beta}\,\d Z^\alpha \d Z^\beta + 2V  k_\alpha k_\beta \,\d Z^\alpha \d Z^\beta,
 \label{KerrSchild}
 \end{equation}
where the first part represents just Minkowski flat space (\ref{MinkowskiD}), $\boldk$ is the null vector defined in section~\ref{trajMink}, and the function $2V$ is
 \begin{equation}
 2V=\frac{2m(u)}{r^{D-3}}+\frac{2\Lambda}{(D-2)(D-1)}\,r^2.
 \label{functionV}
 \end{equation}
The function $m(u)$ corresponds to the decreasing mass of the accelerating object located at ${r=0}$. The associated pure radiation field has the form
 \begin{equation}
 T_{\mu\nu}=\rho\, k_\mu k_\nu\, \quad \hbox{where}\quad  \rho=\frac{n^2(u,\boldn)}{r^{D-2}}\,,
 \label{pureradrhoD}
 \end{equation}
 and
 \begin{equation}
\frac{8\pi}{D-2}\, n^2 =  -m_{,u} + (D-1)\,m\,(\,\log P)_{,u}\,.
 \label{RTequationpureradD}
 \end{equation}
This is obviously a generalisation of expressions (\ref{pureradrho}), (\ref{RTequationpurerad}) valid for ${D=4}$.

\subsection{The Robinson--Trautman form}
\label{Robinson_Trautman}
The same class of exact solutions was independently rediscovered in the context of higher-dimensional Robinson--Trautman spacetimes by Podolsk\'y and Ortaggio~\cite{PodOrt06}. It was demonstrated that in this class the \emph{only} solutions with aligned pure radiation and $\Lambda$ in any dimension ${D>4}$ are of algebraic type~type D and have the form
 \begin{equation}
 \d s^2=\frac{r^2}{P^2}\,\gamma_{ij}\,\d x^i\d x^j-2\,\d u\,\d r-2H\,\d u^2 ,
 \label{geo_metric fin}
 \end{equation}
where the function $2H$ is
 \begin{equation}
 \hspace{-10mm}
2H=\frac{{\cal R}(u)}{(D-2)(D-3)}-2\,r(\,\log P)_{,u}-\frac{2m(u)}{r^{D-3}}-\frac{2\Lambda}{(D-2)(D-1)}\,r^2 .
\label{Hfin}
 \end{equation}
The unimodular metric ${\gamma_{ij}(x^k)}$ and the function ${P(u,x^i)}$ must satisfy the field equations for the transverse spatial metric ${h_{ij}\equiv P^{-2}\gamma_{ij}}$, namely that at any section ${u=u_0=\,}$const. the metric ${h_{ij}(u_0,x^k)}$ describes an Einstein space,
${{\mathcal R}_{ij}=\frac{1}{D-2}{\mathcal R}\,h_{ij} }$, where ${{\mathcal R}_{ij}}$ and ${{\mathcal R}}$ is the corresponding spatial Ricci tensor and scalar, respectively.

The simplest family of such spacetimes arises when the metric $h_{ij}$ is of \emph{constant curvature} and thus conformally flat (this is always true when $D=5$ since $h_{ij}$ is then three-dimensional). In such a case, in suitable coordinates $x^i$,
 \begin{equation}
\gamma_{ij}=\delta_{ij}\,, \label{flat}
 \end{equation}
and, by integrating the remaining field equations, the function $P(u,x^i)$ must have the general form (\ref{RTMinkowski2cartP}), ${P=A(u)+B_i(u)\,x^i+C(u)\,\delta_{ij}\,x^ix^j}$.
In the flat-space limit (${m\to0}$, ${\Lambda\to0}$) the functions $A(u), B_i(u), C(u)$ are related to the components of the velocity $\boldu$ of a test particle at ${r=0}$, as given explicitly by expressions (\ref{RTMinkowskiDabc}). Because this physical interpretation follows from relation (\ref{functionP}), which is the consequence of (\ref{kunormaliz}), (\ref{represk}), and since \begin{equation}
g_{\alpha\beta}\,k^\alpha \dot z^\beta\equiv(\eta_{\alpha\beta}+2V k_\alpha k_\beta)\,k^\alpha \dot z^\beta=\eta_{\alpha\beta}\,k^\alpha \dot z^\beta =-1,
 \label{relationpreserved}
 \end{equation}
such an interpretation of the functions $A(u), B_i(u), C(u)$ remains valid also in the general (non-flat) case ${m\not=0}$. Moreover, due to the coordinate freedom of the metric (\ref{geo_metric fin}), (\ref{Hfin}) given by ${u=u(\tilde u)}$, ${r=\tilde r/u'(\tilde u)}$, where ${u'\equiv\frac{\d u}{\d \tilde u}}$, which implies ${\tilde P=P\,u'}$, ${\tilde {\cal R}={\cal R}\,{u'}^2}$, ${\tilde m=m\,{u'}^{D-1}}$, ${\tilde n^2=n^2\,{u'}^D}$ and thus ${\tilde A=A\,u'}$, ${\tilde B_i=B_i\,u'}$, ${\tilde C=C\,u'}$, the corresponding (positive) Ricci scalar term in (\ref{Hfin}) can always be set equal to 1, see (\ref{RTMinkowski2cartR}). Therefore, the transverse space spanned by $x^i$ is a sphere $S^{D-2}$ of constant positive curvature. The complete form of such Robinson--Trautman pure radiation spacetimes~(\ref{geo_metric fin})--(\ref{flat}), with $m$ and $\Lambda$ non-trivial, thus reads
 \begin{eqnarray}
  \d s^2&=&\frac{r^2}{P^2}\,\delta_{ij}\,\d x^i\d x^j  -2\,\d u\,\d r  \nonumber\\
  &&   -\left(1-2\,r(\,\log P)_{,u}-\frac{2m(u)}{r^{D-3}}-\frac{2\Lambda}{(D-2)(D-1)}\,r^2\right)\d u^2 ,
 \label{geo_metric finF}
 \end{eqnarray}
where the function $P$ is explicitly given by (\ref{functionP2cart}),
\begin{equation}
P = (\dot z^0-\dot z^{D-1})-(\delta_{ij}\,\dot z^j)\,x^i + {\textstyle \frac{1}{4}}(\dot z^0+\dot z^{D-1})\,\delta_{ij}\, x^i x^j\,.
 \label{functionP2cartF}
 \end{equation}
Here ${\dot z^0(u),\dot z^1(u), \ldots , \dot z^{D-1}(u)}$ are components of the velocity (\ref{velocity}) of an arbitrarily moving photon rocket located at ${r=0}$, whose worldline in the background space (\ref{MinkowskiD}) is ${z^\alpha(u)}$. In particular, for an object at rest, $P$ reduces to ${P = 1 + \frac{1}{4}\,\delta_{ij}\, x^i x^j}$ and metric (\ref{geo_metric finF}) is the ${D>4}$ counterpart of the well-known Vaidya--(anti-)de~Sitter metric~\cite{Stephanibook,GriPod09}. Such spacetime is spherically symmetric and radiation of photons is isotropic with ${n^2(u)= -\frac{D-2}{8\pi}\, m_{,u}}$. For constant $m$ we recover vacuum Schwarzschild--Kottler--Tangherlini black hole solution.

It may easily be observed that the metric (\ref{geo_metric finF}) can naturally be decomposed into the flat-space metric (\ref{RTMinkowski2cart}) in the Cartesian-like accelerated coordinates, and the Kerr--Schild term ${ 2V\d u^2 }$, where the function $2V$ is given by (\ref{functionV}). Indeed, for the null vector ${\boldk=k^\alpha\partial_\alpha=\partial_r}$ in the metric (\ref{geo_metric finF}) we obtain ${k_\alpha \,\d Z^\alpha =-\d u}$. This explicitly demonstrates the complete equivalence of the Kerr--Schild form (\ref{KerrSchild}), (\ref{functionV}) and the Robinson--Trautman form (\ref{geo_metric finF}), (\ref{functionP2cartF}) of the metric a photon rocket moving arbitrarily in any dimension ${D\ge4}$.

Notice also that the replacement of the flat background metric $\eta_{\alpha\beta}$ by the curved exact metric ${g_{\alpha\beta}=\eta_{\alpha\beta}+2V k_\alpha k_\beta}$ of the photon-rocket spacetime [(\ref{KerrSchild}) or (\ref{geo_metric finF})] preserves the normalisations (\ref{kunormaliz}) and (\ref{kknormaliz}), namely ${ \boldk\cdot\boldu=g_{\alpha\beta}\,k^\alpha \dot z^\beta =-1}$ and ${\boldk\cdot\boldk=g_{\alpha\beta}\,k^\alpha k^\beta =0}$. On the other hand, the relation (\ref{uunormaliz}) changes to ${\boldu\cdot\boldu=g_{\alpha\beta}\,\dot z^\alpha \dot z^\beta =-(1-2V)}$, where the function ${2V(r,u)}$ is given by expression (\ref{functionV}). The parameter $u$ thus loses its \emph{direct} physical meaning as the proper time of an accelerating test particle located at ${r=0}$. Of course, this is not surprising since ${r=0}$ in the complete metric ${g_{\alpha\beta}}$ corresponds to a curvature singularity, and the redshift factor ${\sqrt{1-2V}=\sqrt{1-2m\,r^{3-D}-\frac{2\Lambda}{(D-2)(D-1)}\,r^2}}$, which describes the time dilation within a gravitational well of the massive photon rocket, also has to be taken into account. In fact, similarly as in figure~\ref{fig1}, the curved spacetime with the metric ${g_{\alpha\beta}}$ is foliated by a family of null hypersurfaces ${u=}$~const., but the coordinate $u$ now plays the role of the time measured by distant observers in  asymptotically flat regions ${r\to\infty}$ where ${V\to0}$ (assuming ${\Lambda=0}$; for the case ${\Lambda\not=0}$ see the following subsection~\ref{cosmolconst}). This justifies the physical interpretation of the functions ${\dot z^\alpha(u)}$ in (\ref{functionP2cartF}) as components of the velocity $\boldu$ of an accelerating rocket, measured with respect to the background flat space (associated with the asymptotic regions far away from the rocket). Moreover, the same interpretation is obtained in a weak-field limit when the mass of the rocket becomes negligible (${m\to0}$).

Finally, it is possible to employ an alternative spherical-like representation of the accelerated background coordinates, namely the flat-space metric form (\ref{RTMinkowskiDspher}), (\ref{functionPDspher}) given in subsection~\ref{trajMinkspher}. The complete spacetime metric ${g_{\alpha\beta}}$ describing a rocket then becomes
 \begin{eqnarray}
  \d s^2&=&{r^2\over p^2}\sum_{i=1}^{D-2} \Big( \prod_{j=1}^{i-1} \sin^2\theta_j \Big) \d\theta_i^2 -2\,\d u\,\d r   \nonumber\\
  &&   -\left(1-2\,r(\,\log p)_{,u}-\frac{2m(u)}{r^{D-3}}-\frac{2\Lambda}{(D-2)(D-1)}\,r^2\right)\d u^2 ,
 \label{RTMinkowskiDspherF}
 \end{eqnarray}
where
\begin{equation}
 p(u,\theta_i)=\dot z^0(u)-\sum_{i=1}^{D-1}\dot z^i(u)\,\cos\theta_i \prod_{j=1}^{i-1} \sin\theta_j \,,
 \label{functionPDspherF}
 \end{equation}
which is fully equivalent to the Cartesian-like form (\ref{geo_metric finF}), (\ref{functionP2cartF}).

\subsection{The inclusion of a cosmological constant}
\label{cosmolconst}
In the above metrics, the cosmological constant $\Lambda$ may take an arbitrary value. Therefore, the solutions represent a photon rocket moving not only in a $D$-dimensional (asymptotically) Minkowski space (when ${\Lambda=0}$), but also in de~Sitter (${\Lambda>0}$) or anti-de~Sitter universe (${\Lambda<0}$). These are the three maximally symmetric, conformally flat spacetimes of constant curvature ${R=\frac{2D}{D-2}\Lambda}$.

For ${\Lambda\not=0}$, the functions ${\dot z^0(u),\dot z^1(u), \ldots , \dot z^{D-1}(u)}$ in (\ref{functionP2cartF}) or (\ref{functionPDspherF}) still retain their physical meaning as components of the velocity of the photon rocket with respect to the ``background'' frame ${\partial_\alpha}$ corresponding to the coordinates $Z^\alpha$ of (\ref{MinkowskiD}). However, these now need to be understood as coordinates of an ``external'' flat space of dimension ${D+1}$, into which the (anti-)de~Sitter universe is embedded.

To be specific, it is well known that the de~Sitter and anti-de~Sitter spacetimes can be represented as a $D$-dimensional hyperboloid
 \begin{equation}
 -(Z^0)^2+(Z^1)^2+\cdots +(Z^{D-1})^2+\epsilon\,(Z^D)^2 = \frac{(D-2)(D-1)}{2\Lambda} \,,
 \label{hyperboloid}
 \end{equation}
embedded in a ${(D+1)}$-dimensional flat space with the metric
\begin{equation}
 \d s^2=-{(\d Z^0)}^2+{(\d Z^1)}^2+\cdots +{(\d Z^{D-1})}^2+\epsilon\,{(\d Z^D)}^2 ,
 \label{MinkowskiDplus1}
 \end{equation}
where ${\,\epsilon\equiv\hbox{sign}\,\Lambda\,}$.

As in (\ref{worldline}), we may now consider a timelike worldline ${z^\alpha(u)}$ in this ${(D+1)}$-dimensional flat space.
If we \emph{assume} that the functions ${z^\alpha(u)}$ satisfy the constraint
 \begin{equation}
\hspace{-20mm}
 -[z^0(u)]^2+[z^1(u)]^2+\cdots +[z^{D-1}(u)]^2+\epsilon\,[z^D(u)]^2 = \frac{(D-2)(D-1)}{2\Lambda}
 \label{hyperboloidtraject}
 \end{equation}
at any time $u$, the photon rocket (with a negligible mass $m$) during its motion will always remain on the hyperboloid (\ref{hyperboloid}), i.e., in the $D$-dimensional (anti-)de~Sitter universe. The corresponding ${(D+1)}$-velocity ${\boldu=\dot z^\alpha(u)\,\partial_\alpha}$ is normalised to
 \begin{equation}
 -[\dot z^0(u)]^2+[\dot z^1(u)]^2+\cdots +[\dot z^{D-1}(u)]^2+\epsilon\,[\dot z^D(u)]^2 = -1\,,
 \label{constvelosD}
 \end{equation}
(for ${\Lambda<0}$ we simply modify the metric $\eta_{\alpha\beta}$ to ${\hbox{diag}(-1,+1,\cdots,+1,-1)}$). This constraint may be considered as the relation which \emph{determines} the value of the function $\dot z^D(u)$ in terms of the velocity components ${\dot z^0(u),\dot z^1(u), \ldots , \dot z^{D-1}(u)}$ which occur in the functions (\ref{functionP2cartF}) or (\ref{functionPDspherF}). Effectively, it thus only remains to choose the trajectory of the rocket in such a way that it satisfies the condition (\ref{hyperboloidtraject}).

Notice finally that if the mass of the rocket vanishes, the metric (\ref{geo_metric finF}) or (\ref{RTMinkowskiDspherF}) with ${m=0}$ and ${\Lambda\not=0}$ is just the de~Sitter or anti-de~Sitter $D$-dimensional space, expressed in coordinates whose origin ${r=0}$ is arbitrarily accelerating. An explicit representation of the corresponding hyperboloids (\ref{hyperboloid}) in terms of these coordinates can be obtained using the transformations (\ref{transformation}) and (\ref{represk}) as
 \begin{eqnarray}
 Z^0&=&z^0(u)+ \frac{r}{p(u,\boldn)}\,, \nonumber\\
 Z^i&=&z^i(u)+ \frac{r}{p(u,\boldn)}\,n^i\,,  \quad i=1,\ldots,D\,,\label{transformationsinglefff}
  \end{eqnarray}
in which either (\ref{nicartD}) or (\ref{nispherD}) is employed to parameterise the components $n^i$ of the unit vector $\boldn$. For the case ${D=4}$ such relations were recently discussed in detail in \cite{Pod08}.

\section{Motion of the rocket in a single direction}
\label{singlesection}

Let us now concentrate on a particular case in which the photon rocket in $D$ dimensions  accelerates always in the same spatial direction, i.e., without performing any manoeuvres. It may only change its speed. In such a situation, without loss of generality it is possible to rotate the spatial axes of the background flat space so that the timelike worldline (\ref{worldline}) of the rocket is
 \begin{equation}
 z^\alpha(u)=\{z^0(u),z^1(u),0,0, \ldots, 0\} \,.
 \label{worldlinesingle}
 \end{equation}
The corresponding velocity (\ref{velocity}) is ${\boldu=\{\dot z^0,\dot z^1,0,0, \ldots, 0\}}$ and the condition (\ref{uunormaliz}) reduces to
 \begin{equation}
 (\dot z^0)^2-(\dot z^1)^2=1\,,
  \label{velocitysingle}
 \end{equation}
The metric of the photon rocket, expressed in the spherical-like coordinates (\ref{RTMinkowskiDspherF}), (\ref{functionPDspherF}), thus becomes
 \begin{eqnarray}
  \d s^2&=&{r^2\over p^2}\sum_{i=1}^{D-2} \Big( \prod_{j=1}^{i-1} \sin^2\theta_j \Big) \d\theta_i^2 - 2\,\d u\,\d r   \nonumber\\
  &&   -\left(1-2\,r(\,\log p)_{,u}-\frac{2m(u)}{r^{D-3}}-\frac{2\Lambda}{(D-2)(D-1)}\,r^2\right)\d u^2 ,
 \label{RTMinkowskiDsphersingle}
 \end{eqnarray}
where
 \begin{equation}
 p(u,\theta_1)=\dot z^0(u)-\dot z^1(u)\,\cos\theta_1  \,.
 \label{functionPDsphersingle}
 \end{equation}

In the test-particle limit when the mass of the rocket is negligible (${m\to0}$) and ${\Lambda=0}$, the background flat-space coordinates $Z^\alpha$ are obtained by the transformation (\ref{transformation}). Using the relations (\ref{represk}) and (\ref{nispherD}) this explicitly becomes
 \begin{eqnarray}
 Z^0&=&z^0(u)+ \frac{r}{p(u,\theta_1)}\,, \nonumber\\
 Z^1&=&z^1(u)+ \frac{r}{p(u,\theta_1)}\,\cos\theta_1 \,,  \label{transformationsingle}\\
 Z^2&=& \hspace{12.7mm}  \frac{r}{p(u,\theta_1)}\,\sin\theta_1\,\cos\theta_2\,, \nonumber\\
 Z^i&=& \hspace{12.7mm}  \frac{r}{p(u,\theta_1)}\,\sin\theta_1\Big(\prod_{j=2}^{i-1} \sin\theta_j\Big)\cos\theta_i\,, \quad i=3,\ldots,D-1\,. \nonumber
 \end{eqnarray}
where, as in (\ref{nispherDdef}), we define ${\theta_{D-1}\equiv0}$. Obviously, at any time the origin ${r=0}$ of the accelerated coordinates of (\ref{RTMinkowskiDsphersingle}) coincides with the rocket.

It now appears to be convenient to introduce the following parametrisation of the components of the velocity $\boldu$:
 \begin{eqnarray}
 \dot z^0 &=& \cosh \big({\textstyle \int}\alpha(u)\,\d u\big)\,, \nonumber\\
 \dot z^1 &=& \sinh \big({\textstyle \int}\alpha(u)\,\d u\big)\,,  \label{accelparam}
 \end{eqnarray}
which identically satisfies the condition (\ref{velocitysingle}). The acceleration vector $\bolda$ has the only nonvanishing components
${\ddot z^{\,0}= \alpha(u)\,\dot z^1}$, ${\ddot z^{\,1}= \alpha(u)\,\dot z^0}$, so that $\bolda\cdot\boldu=0$. Moreover, $\bolda\cdot\bolda=\alpha^2(u)$ which means that the function $\alpha(u)$ is exactly the value of \emph{instantaneous acceleration of the rocket} at the time $u$. Using this parametrisation, the metric function (\ref{functionPDsphersingle}) becomes
 \begin{equation}
 p(u,\theta_1)=\cosh \big({\textstyle \int}\alpha(u)\,\d u\big) - \cos\theta_1\,\sinh \big({\textstyle \int}\alpha(u)\,\d u\big)  \,.
 \label{functionPDsphersinglea}
 \end{equation}
Finally, it is possible to define a new angular coordinate $\vartheta$ as
 \begin{equation}
 \sin\vartheta \equiv \frac{\sin\theta_1}{p(u,\theta_1)} \,.
 \label{newtheta}
 \end{equation}
Straightforward calculation shows that (\ref{newtheta}) implies the following interesting identities:
 \begin{eqnarray}
 \sin\vartheta &=& \frac{\sin\theta_1}
   {\cosh \big({\textstyle \int}\alpha(u)\,\d u\big) - \cos\theta_1\,\sinh \big({\textstyle \int}\alpha(u)\,\d u\big) }\,, \nonumber\\
 \cos\vartheta &=& \frac{\sinh \big({\textstyle \int}\alpha(u)\,\d u\big) - \cos\theta_1\,\cosh \big({\textstyle \int}\alpha(u)\,\d u\big)}
   {\cosh \big({\textstyle \int}\alpha(u)\,\d u\big) - \cos\theta_1\,\sinh \big({\textstyle \int}\alpha(u)\,\d u\big) }\,, \nonumber\\
 \cot\frac{\theta_1}{2} &=& \tan\frac{\vartheta}{2} \,\exp\big({{\textstyle \int}\alpha(u)\,\d u}\big) \,, \label{identities}\\
 (\,\log p)_{,u}&=& \alpha(u)\, \cos\vartheta\,,\nonumber\\
   p^{-1} &=& \cosh \big({\textstyle \int}\alpha(u)\,\d u\big) - \cos\vartheta\,\sinh \big({\textstyle \int}\alpha(u)\,\d u\big) \,, \nonumber\\
 p^{-1}\cos\theta_1  &=& \sinh \big({\textstyle \int}\alpha(u)\,\d u\big) - \cos\vartheta\,\cosh \big({\textstyle \int}\alpha(u)\,\d u\big) \,, \nonumber\\
 &&\hspace{-19.5mm}  p^{-2} (\d \theta_1 ^2 + \sin^2\theta_1 \,\d \theta_2^2 ) = (\d \vartheta + \alpha\,\sin\vartheta\,\d u)^2+\sin^2\vartheta \,\d \,\theta_2^2 \,. \nonumber
 \end{eqnarray}
With these relations, the parametrisation (\ref{transformationsingle}) of the Minkowski background (when ${m=0=\Lambda}$) in terms of the accelerated coordinates becomes
 \begin{eqnarray}
 Z^0&=&z^0(u)+ r\left[\,\cosh \big({\textstyle \int}\alpha(u)\,\d u\big) - \cos\vartheta\,\sinh \big({\textstyle \int}\alpha(u)\,\d u\big)\,\right], \nonumber\\
 Z^1&=&z^1(u)+ r\left[\,\sinh \big({\textstyle \int}\alpha(u)\,\d u\big) - \cos\vartheta\,\cosh \big({\textstyle \int}\alpha(u)\,\d u\big)\,\right],  \label{transformationsinglealter}\\
 Z^2&=& \hspace{12.7mm}  r\, \sin\vartheta\,\cos\theta_2\,, \nonumber\\
 Z^i&=& \hspace{12.7mm}  r\, \sin\vartheta\,\Big(\prod_{j=2}^{i-1}  \sin\theta_j\Big)\cos\theta_i\,, \quad i=3,\ldots,D-1\,. \nonumber
 \end{eqnarray}
The complete metric (\ref{RTMinkowskiDsphersingle}) takes the form
\begin{eqnarray}
&& \,\hspace{-22mm}\d s^2 = r^2 \big(\d \vartheta + \alpha(u)\,\sin\vartheta\,\d u\big)^2+r^2\sin^2\vartheta \,\Big( \d \,\theta_2^2 + \sum_{i=3}^{D-2} \Big( \prod_{j=2}^{i-1}  \sin^2\theta_j \Big) \d\theta_i^2 \Big)  \nonumber\\
 && \hspace{-12mm}  -2\,\d u\,\d r  - \Big(1-2\,r\,\alpha(u)\, \cos\vartheta-\frac{2m(u)}{r^{D-3}}-\frac{2\Lambda}{(D-2)(D-1)}\,r^2 \Big)\,\d u^2,
 \label{RTMinkowskiDsphersinglealter}
 \end{eqnarray}
which can be rewritten as
 \begin{eqnarray}
&& \,\hspace{-22mm}  \d s^2 = -\Big(1-\frac{2m(u)}{r^{D-3}}-\frac{2\Lambda}{(D-2)(D-1)}\,r^2-2\,\alpha(u)\,r\,\cos\vartheta -\alpha^2(u)\,r^2\sin^2\vartheta \Big)\,\d u^2   \nonumber\\
 && \hspace{-12mm}    -2\,\d u\,\d r +\>2\,\alpha(u)\,r^2\sin\vartheta\,\d u\,\d\vartheta  \nonumber\\
 && \hspace{-12mm}    + r^2 \bigg(\d \vartheta^2 +\sin^2\vartheta \Big( \d \,\theta_2^2 + \sum_{i=3}^{D-2} \Big( \prod_{j=2}^{i-1} \sin^2\theta_j \Big) \d\theta_i^2 \Big)\bigg).  \label{RTMinkowskiDsphersinglealter2}
 \end{eqnarray}
This is a generalisation, to any ${D\ge4}$, of the standard metric form (\ref{Kinnersrocketmetric}) of the Kinnersley photon rocket \cite{Kin69,Bon94,vonGonKram98,Pod08} accelerating arbitrarily in a single spatial direction. In the absence of acceleration, ${\alpha(u)=0}$, it reduces to the higher-dimensional Vaidya--(anti-)de~Sitter  solution which describes a spherically symmetric fixed source at ${r=0}$ with a varying mass determined by $m(u)$.

Alternatively, the metric for the photon rocket accelerating in a single spatial direction (now $\partial_{D-1}$) can be written in the Cartesian-like coordinates (\ref{geo_metric finF}), (\ref{functionP2cartF}) with $P$ simplified to ${P = (\dot z^0-\dot z^{D-1}) + {\textstyle \frac{1}{4}}(\dot z^0+\dot z^{D-1})\,\delta_{ij}\, x^i x^j}$. Following (\ref{accelparam}), this becomes
\begin{equation}
P = \exp \big(\!-{\textstyle \int}\alpha(u)\,\d u\big) + {\textstyle \frac{1}{4}}\exp \big({\textstyle \int}\alpha(u)\,\d u\big)\,\delta_{ij}\, x^i x^j\,.
 \label{functionP2cartsingle}
 \end{equation}

\section{Explicit examples of motion of a photon rocket}
\label{examplsection}

It will now be illustrative to investigate in more detail some particular examples which describe accelerating or decelerating photon rockets along straight or curved trajectories. In all the cases we will identify and study the particular situation in which the minimal amount of radiation is emitted, i.e., the mass decrease of the rocket (its ``energy consumption'') for a given flight is minimised.

\subsection{Straight flight}
\label{singlesectionstraight}

The simplest situation arises when the photon rocket moves in a single spatial direction. It does not perform any manoeuvres, except for changing its speed.  As described in section~\ref{singlesection}, in such a case the timelike worldline $z^\alpha(u)$ of the rocket is fully described by (\ref{worldlinesingle}), i.e., by the two functions ${z^0(u)}$ and ${z^1(u)}$. With the parametrisation of the corresponding velocity $\boldu$ given by (\ref{accelparam}),
 \begin{equation}
 \dot z^0(u) = \cosh \big({\textstyle \int}\alpha(u)\,\d u\big)\,, \qquad
 \dot z^1(u) = \sinh \big({\textstyle \int}\alpha(u)\,\d u\big)\,,  \label{accelparamstraight}
 \end{equation}
where $\alpha(u)$ is the acceleration of the rocket as a function of the time $u$, the metric takes the form (\ref{RTMinkowskiDsphersinglealter2}). It follows from relations (\ref{transformationsinglealter}) that the angle $\vartheta$, introduced in (\ref{newtheta}), takes the value ${\vartheta=0}$ \emph{behind} the rocket while ${\vartheta=\pi}$ \emph{in front of} the rocket (considering ${\alpha>0}$).

\begin{figure}[ht]
   \vspace{-2mm}
   \begin{center}
   \includegraphics[scale=0.75]{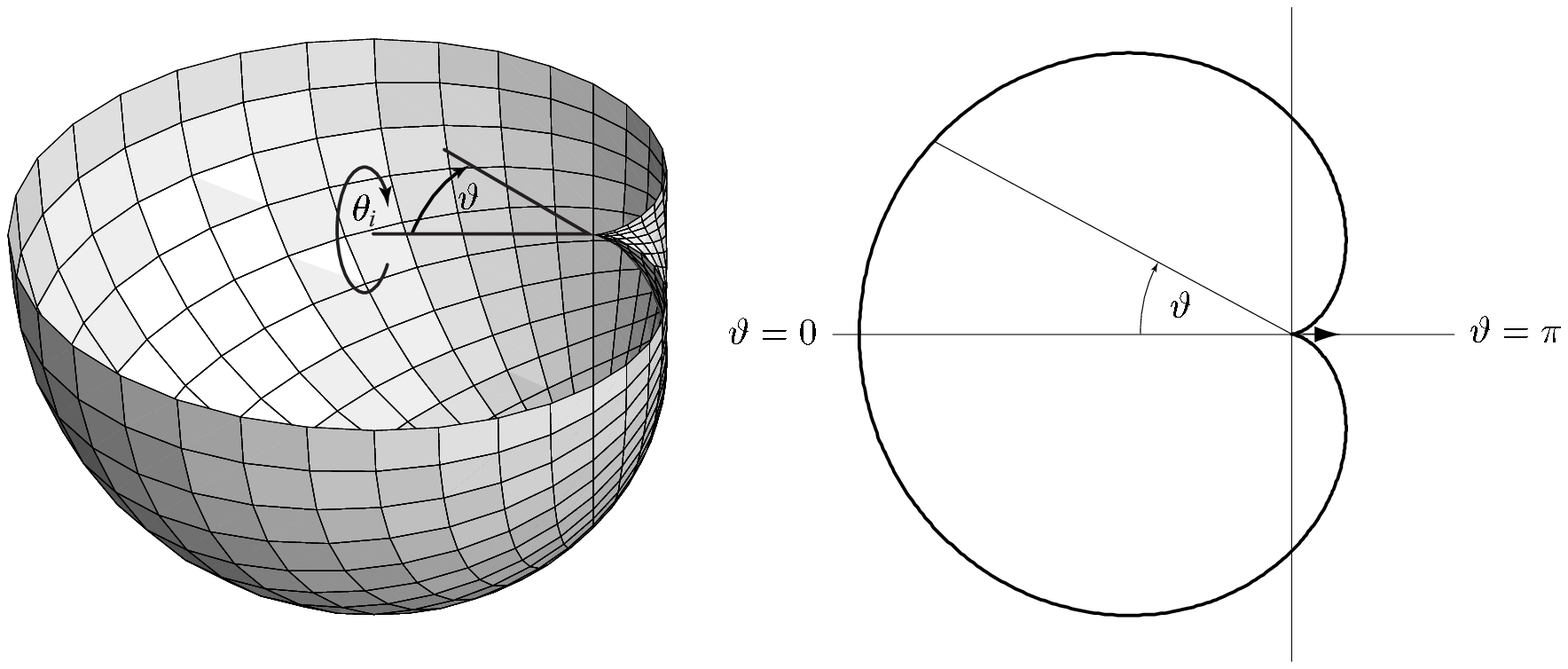}
   \end{center}
   \vspace{-5mm}
   \caption{\label{fig2}%
   Section through the radiation pattern ${n^2}$ of the photon field (\ref{pureradoptimal}) emitted by the rocket which accelerates along a straight line (moving here to the right), as described by the axisymmetric metric (\ref{RTMinkowskiDsphersinglealter2}). On the right we plot the same pattern with the trivial angular coordinates $\theta_i$ (${i=2,\ldots,D-2}$) suppressed.}
   \end{figure}

The associated field of emitted photons has the form (\ref{pureradrhoD}) where the profile $n$ is given by (\ref{RTequationpureradD}). Using the relation ${(\,\log P)_{,u}=(\,\log p)_{,u}=\alpha(u)\,\cos\vartheta}$, see (\ref{relationPp}) and (\ref{identities}),
 \begin{equation}
 n^2(u,\vartheta) = \frac{D-2}{8\pi}\,\big[   -m_{,u}(u) + (D-1)\,\alpha(u)\,m(u)\cos\vartheta\,\big].
 \label{RTequationpureradDexmpl}
 \end{equation}
Obviously, the maximum of radiation is emitted directly behind the rocket (for ${\vartheta=0}$) and minimum in the direction in front of it (for ${\vartheta=\pi}$). In fact, it is possible to rewrite the expression (\ref{RTequationpureradDexmpl}) as
 \begin{equation}
 \hspace{-20mm}
 n^2(u,\vartheta) =  \frac{(D-1)(D-2)}{8\pi}\,\alpha\,m\,(1+ \cos\vartheta\,)-\frac{D-2}{8\pi}\,\big[m_{,u}+(D-1)\,\alpha\,m\,\big].
 \label{RTequationpureradDexmplmod}
 \end{equation}
The first term is always positive, and vanishes for ${\vartheta=\pi}$. Therefore, the second term must also be nonnegative. An optimised situation occurs when the second term vanishes: \emph{in such a case the rocket emits no photons in the direction of its motion}, it only emits ``backwards''. This condition yields the explicit relation
 \begin{equation}
 m(u) = m_0\, \exp\big({\textstyle -(D-1)\int}\alpha(u)\,\d u\big).
 \label{moptimal}
 \end{equation}
 The mass of such photon rocket decreases exponentially from its initial value $m_0$, and the corresponding radiation pattern (\ref{RTequationpureradDexmplmod}) is explicitly given as
  \begin{equation}
 n^2(u,\vartheta) =  \frac{D-2}{4\pi}\big(- m_{,u}\big)\cos^2\frac{\vartheta}{2}\,,
 \label{pureradoptimal}
 \end{equation}
 where
  \begin{equation}
 - m_{,u}(u)=  (D-1)\, m_0\,\alpha(u) \exp\Big({\textstyle -(D-1)\int}\alpha(u)\,\d u\Big).
 \label{pureradoptimalB}
 \end{equation}
At a given time $u$, the angular dependence of the photon field on $\vartheta$ and $\theta_i$ is plotted as a spherical polar diagram in figure~\ref{fig2}.

\begin{figure}[ht]
   \vspace{-2mm}
   \begin{center}
   \includegraphics[scale=0.70]{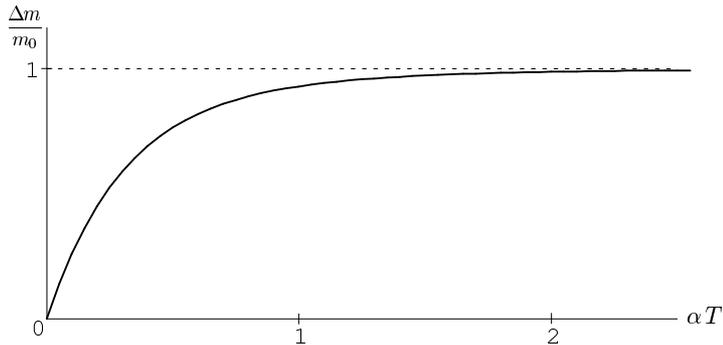}
   \end{center}
   \vspace{-5mm}
   \caption{\label{fig3}%
   The function ${\,\frac{\Delta m}{m_0}(\alpha\, T)\,}$ shows the relative mass decrease of the rocket with the inertial time measured ``on the Earth''.}
   \end{figure}

In particular, for a photon rocket which moves with a \emph{constant acceleration\,}  ${\alpha=}$~const., it is possible to calculate its motion and the total decrease of its mass explicitly. Integrating (\ref{accelparamstraight}) for the initial conditions ${z^0(0)=0=z^1(0)}$ we obtain
 \begin{equation}
 z^0(u) = \frac{1}{\alpha}\sinh (\alpha\, u)\,, \qquad
 z^1(u) = \frac{1}{\alpha}\big[\cosh (\alpha\, u)-1\big]\,.  \label{motionstraight}
 \end{equation}
In view of the transformation (\ref{transformationsingle}), these two functions give exactly the time elapsed $T\equiv Z^0=z^0(u)$ and the distance ${L\equiv Z^1=z^1(u)}$ of the rocket (located at ${r=0}$) with respect to the background inertial frame connected to ``the Earth''. Recall that $u$ is related to the proper time of the photon rocket. We have thus recovered the well-known relations for (integrated) time dilation and length contraction which are valid for a uniformly accelerated objects in special relativity. The corresponding total decrease of the mass of the rocket is given by
 \begin{equation}
 \Delta m \equiv m_0-m(u) = m_0 \big(1-e^{-(D-1)\,\alpha\, u}\big),  \label{massdescreasestraight}
 \end{equation}
where, considering (\ref{motionstraight}),
 \begin{equation}
 \alpha\, u = \hbox{arcsinh}\, (\alpha\, T)= \hbox{arccosh}\, (1+\alpha\, L)\,.  \label{alphaustraight}
 \end{equation}
 The plot of ${\,\Delta m/m_0\,}$ as a function of ${\,\alpha\, T\,}$ for ${D=4}$ is shown in figure~\ref{fig3}.

\noindent
Asymptotically, for large values of $u$, $T$ and $L$, the function (\ref{massdescreasestraight}) reduces to
 \begin{equation}
 \frac{\Delta m}{m_0} \approx  1-(2\alpha\, T)^{1-D} \approx 1-(2\alpha\, L)^{1-D}\,,  \label{massdescreasestraightas}
 \end{equation}
 which demonstrates that the total loss of mass ${\Delta m}$ approaches the initial mass $m_0$ as $T^p$ and $L^p$, where ${p\equiv1-D<0}$. For higher dimensions $D$, the mass decrease is faster, which makes the travel of photon rockets in higher dimensions more demanding.

It is also useful to express the mass function (\ref{massdescreasestraight}) of the rocket in terms of its actual speed $v$ with respect to the inertial frame, ${v=\dot z^1/\dot z^0=\tanh(\alpha\, u)}$ which implies $e^{\alpha\, u}=\sqrt{(1+v)/(1-v)}\,$:
 \begin{equation}
 \frac{m(v)}{m_0} = \left(\frac{1-v}{1+v}\right)^{(D-1)/2}\,.  \label{massdescreasestraightveloc}
 \end{equation}
This simple formula gives the mass ${m(v)}$ of the photon rocket after it was uniformly accelerated from rest (${v=0}$) to the speed $v$.

Another explicitly solvable model of a photon rocket moving in a single direction arises if the \emph{acceleration function} takes the form  ${\,\alpha(u)=\tanh u\,}$ for ${u\ge0}$. In such a case, which may describe specific initial phase of acceleration of the rocket, the relation (\ref{accelparamstraight}) can be integrated to
 \begin{equation}
 \dot z^0(u) = \frac{1}{2}\Big(\!\cosh u + \frac{1}{\cosh u}\, \Big), \quad
 \dot z^1(u) = \frac{1}{2}\Big(\!\cosh u - \frac{1}{\cosh u}\, \Big),  \label{motionstraight2}
 \end{equation}
so that
\begin{eqnarray}
  && T = z^0(u) = \frac{1}{2}\sinh u + \hbox{arctan\,} e^u-\frac{\pi}{4}\,, \nonumber\\
  && L = z^1(u) = \frac{1}{2}\sinh u - \hbox{arctan\,} e^u+\frac{\pi}{4}\,.  \label{motionstraight3}
 \end{eqnarray}
Compared to the case of a uniform acceleration ${\alpha=}$~const., as described by (\ref{motionstraight}), it follows that for large values of the time $u$ the inertial time $T$ and the distance $L$ traveled are asymptotically \emph{half} of those corresponding to ${\alpha=1}$.
The total loss of mass of the rocket is now
 \begin{equation}
 \Delta m = m_0 \big(1-\cosh^{1-D}\!u\big),  \label{massdescreasestraight2}
 \end{equation}
which has the same asymptotic behaviour as (\ref{massdescreasestraightas}) for ${\alpha=1}$.

\subsection{Circular trajectory}
\label{circular}

Let us also study special non-geodetic flight of the photon rocket, namely a circular motion. We assume that the rocket in $D$~dimensions moves along a circle of radius $a$ with a constant angular velocity $\omega$,
 \begin{eqnarray}
  && z^0(u) = \sqrt{1+a^2\omega^2}\,u\,, \nonumber\\
  && z^1(u) = \ldots = z^{D-3}(u) = 0\,,   \nonumber\\
  && z^{D-2}(u) = a\,\cos(\omega\,u)\,, \nonumber\\
  && z^{D-1}(u) = a\,\sin(\omega\,u)\,.
 \label{circtraj}
 \end{eqnarray}
Here $\omega$ is the angular velocity of a photon rocket with respect to the time~$u$. Due to the time dilation/length contraction, the constant speed~$v$ of the rocket on its circular motion, measured with respect to the inertial background frame in the center, is not ${a\omega}$ but
 \begin{equation}
  v=\frac{a\omega}{\sqrt{1+a^2\omega^2}}<1\,, \qquad \hbox{i.e.,} \qquad a\omega=\frac{v}{\sqrt{1-v^2}}\,.
  \label{circvelocity}
 \end{equation}
The velocity parameter ${a\omega}$ may thus take \emph{any value}, and ${v \to 1}$ as ${a\omega \to \infty}$.

Explicit exact metric which describes such motion of the photon rocket is
\begin{eqnarray}
 \d s^2&=&\frac{r^2}{p^2}\,\sum_{i=1}^{D-2} \Big( \prod_{j=1}^{i-1} \sin^2\theta_j \Big) \d\theta_i^2 - 2\,\d u\,\d r   \nonumber\\
  &&   -\left(1-2\,r(\,\log p)_{,u}-\frac{2m(u)}{r^{D-3}}-\frac{2\Lambda}{(D-2)(D-1)}\,r^2\right)\d u^2 ,
 \label{circmetr}
\end{eqnarray}
where \begin{equation}
 p(u,\theta_j,\phi)=\sqrt{1+a^2\omega^2}-a\omega\,\Big( \prod_{j=1}^{D-3} \sin\theta_j \Big)\sin(\phi-\omega\,u)\,,
 \label{circfunction}
 \end{equation}
${\theta_1, \theta_2,\ldots,\theta_{D-3}\in[0,\pi]}$ and ${\phi\equiv\theta_{D-2}\in[0,2\pi)}$, see (\ref{RTMinkowskiDspherF}) and (\ref{functionPDspherF}).

The corresponding radiation pattern of emitted photons is given by (\ref{RTequationpureradD}). Since $(\,\log P)_{,u}=(\,\log p)_{,u}\,$, see (\ref{relationPp}), this becomes
 \begin{equation}
  n^2(u,\theta_j,\phi) = \frac{D-2}{8\pi}\big[-m_{,u} + (D-1)\,a\omega^2m\,f(u,\theta_j,\phi)\big],
  \label{circradD}
 \end{equation}
in which we introduced
 \begin{equation}
  f(u,\theta_j,\phi) \equiv \frac{\Big( \prod_{j=1}^{D-3} \sin\theta_j \Big)\cos(\phi-\omega\,u)}{\sqrt{1+a^2\omega^2}-a\omega\Big( \prod_{j=1}^{D-3} \sin\theta_j \Big)\sin(\phi-\omega\,u)}\,.
  \label{circf}
 \end{equation}
In ${D=4}$ this simplifies to ${\Big( \prod_{j=1}^{D-3} \sin\theta_j \Big)=\sin\theta_1\equiv\sin\theta}$, and the angular part of the metric (\ref{circmetr}) reduces to ${(\d \theta^2+\sin^2\,\theta\, \d \phi^2)}$.

\begin{figure}[ht]
   \vspace{-2mm}
   \begin{center}
   \includegraphics[scale=0.70]{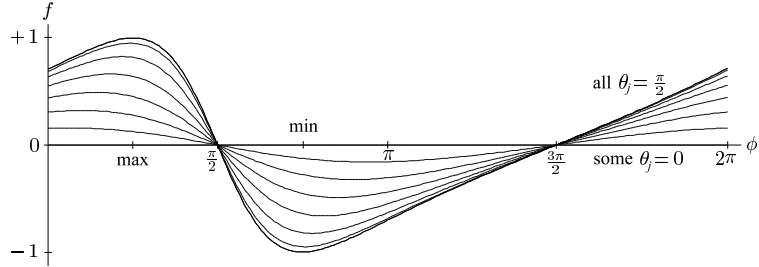}
   \end{center}
   \vspace{-5mm}
   \caption{\label{fig4}%
   The function ${f(u,\theta_j,\phi)}$ for ${u=0}$ and several constant values of $\theta_j$ between $0$ and~${\frac{\pi}{2}}$. Here we set ${a=1=\omega}$.}
   \end{figure}

The function~$f$ identically vanishes whenever ${\theta_j=0}$ and ${\theta_j=\pi}$ which, due to the definition (\ref{nispherD}), describe directions perpendicular to the plane of the circular trajectory (\ref{circtraj}) of the photon rocket. Also, ${f=0}$ if ${\phi-\omega\,u=\frac{\pi}{2}}$ or ${\phi-\omega\,u=\frac{3\pi}{2}}$. For fixed~$\phi$ and~$u$, the maximum of the function $f$ occurs at ${\theta_j=\frac{\pi}{2}}$ for all ${j=1, \ldots, D-3}$, that is in the plane of motion. If ${\theta_j=\frac{\pi}{2}}$ for all $j$, and $u$ is fixed, the extremes of $f$ are given by the condition ${\sin(\phi-\omega\,u)=a\omega/\sqrt{1+a^2\omega^2}}$ which implies ${\,\tan(\phi-\omega\,u)=\pm a\omega}$. For ${\phi-\omega\,u<\frac{\pi}{2}}$ the upper sign applies and the corresponding direction gives the \emph{maximal} value of $f$, namely ${f=+1}$. Contrary, for ${\phi-\omega\,u>\frac{\pi}{2}}$ the lower sign gives the direction in which there is a \emph{minimal} value, ${f=-1}$, in the radiation pattern. A typical plot of the function ${f(u=0,\theta_j,\phi)}$ for several constant values of $\theta_j$ is shown in figure~\ref{fig4}.

\noindent
Since ${f(u,\theta_j,\phi)\in[-1,+1]}$ and the minimal value ${f=-1}$ is attained, to satisfy the condition ${n^2(u,\theta_j,\phi)\ge0}$ everywhere, we rewrite (\ref{circradD}) as
 \begin{equation}
  n^2(u,\theta_j,\phi) = \frac{(D-1)(D-2)}{8\pi}a\omega^2m\big[1 + f(u,\theta_j,\phi)\big]-\frac{D-2}{8\pi}\big[m_{,u}+(D-1)\,a\omega^2m\big] .
  \label{circradDrew}
 \end{equation}
Similarly as in (\ref{RTequationpureradDexmplmod}), the first term is nonnegative and vanishes at the minimum of $f$. An optimised flight of the photon rocket, which minimises the mass decrease, thus occurs if the second term in (\ref{circradDrew}) vanishes, i.e., when
 \begin{equation}
 m(u) = m_0\, \exp\big( -(D-1)\,a\omega^2\, u\big).
 \label{circmoptimal}
 \end{equation}
The mass of the photon rocket then decreases exponentially. The coefficient ${a\omega^2}$ is, in fact, the classical centrifugal acceleration (notice the analogy with expression (\ref{moptimal}) for a straight trajectory, in particular when the acceleration $\alpha$ is constant).  The corresponding radiation pattern is
  \begin{equation}
 n^2(u,\theta_j,\phi) =  \frac{D-2}{8\pi}\big(- m_{,u}\big)\big[1 + f(u,\theta_j,\phi)\big].
 \label{circpureradoptimal}
 \end{equation}

\begin{figure}[ht]
   \vspace{-6mm}
   \begin{center}
   \includegraphics[scale=0.8]{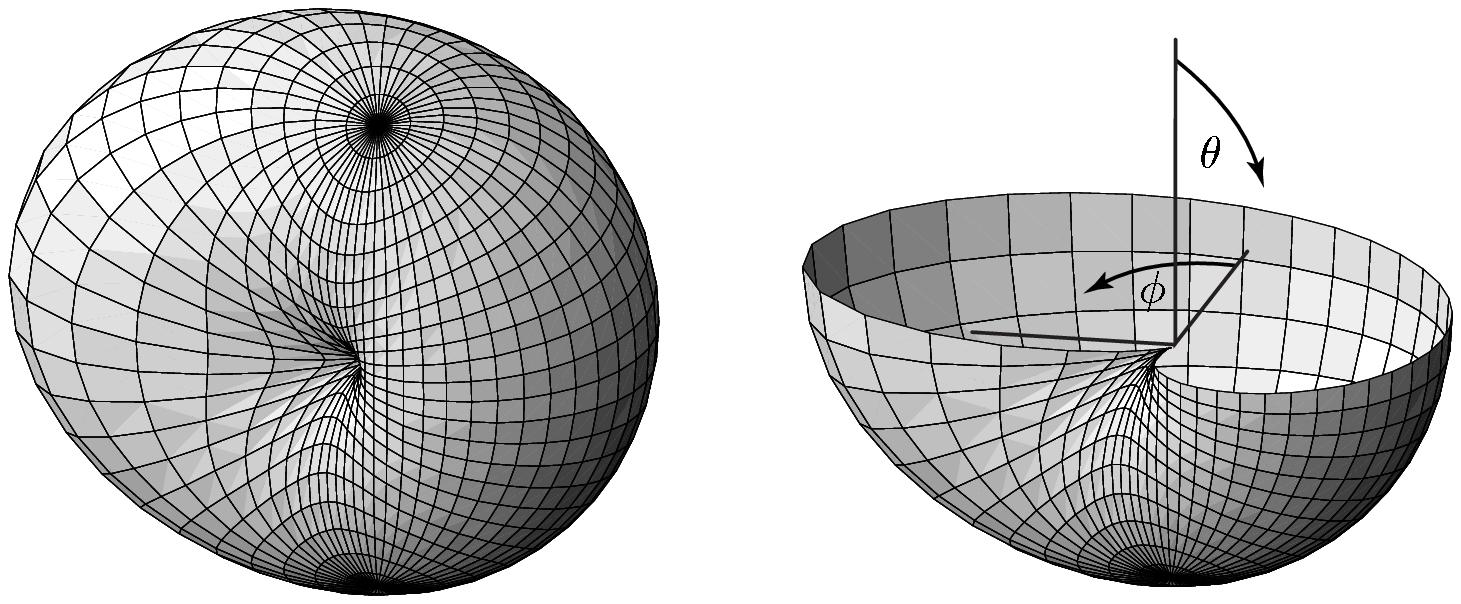}
   \end{center}
   \vspace{-5mm}
   \caption{\label{fig5}%
   Radiation pattern ${n^2(\theta,\phi)}$ of the photon field (\ref{circpureradoptimal}) emitted (at ${u=0}$) by the rocket moving in four dimensions along  a circle of radius~${a=1}$ with the angular velocity~${\omega=1}$ (left). Equatorial section ${\theta=\frac{\pi}{2}}$ through the diagram, where ${n^2(\theta,\phi)}$ reaches extreme values (right).}
   \vspace{-2mm}
   \begin{center}
   \includegraphics[scale=0.8]{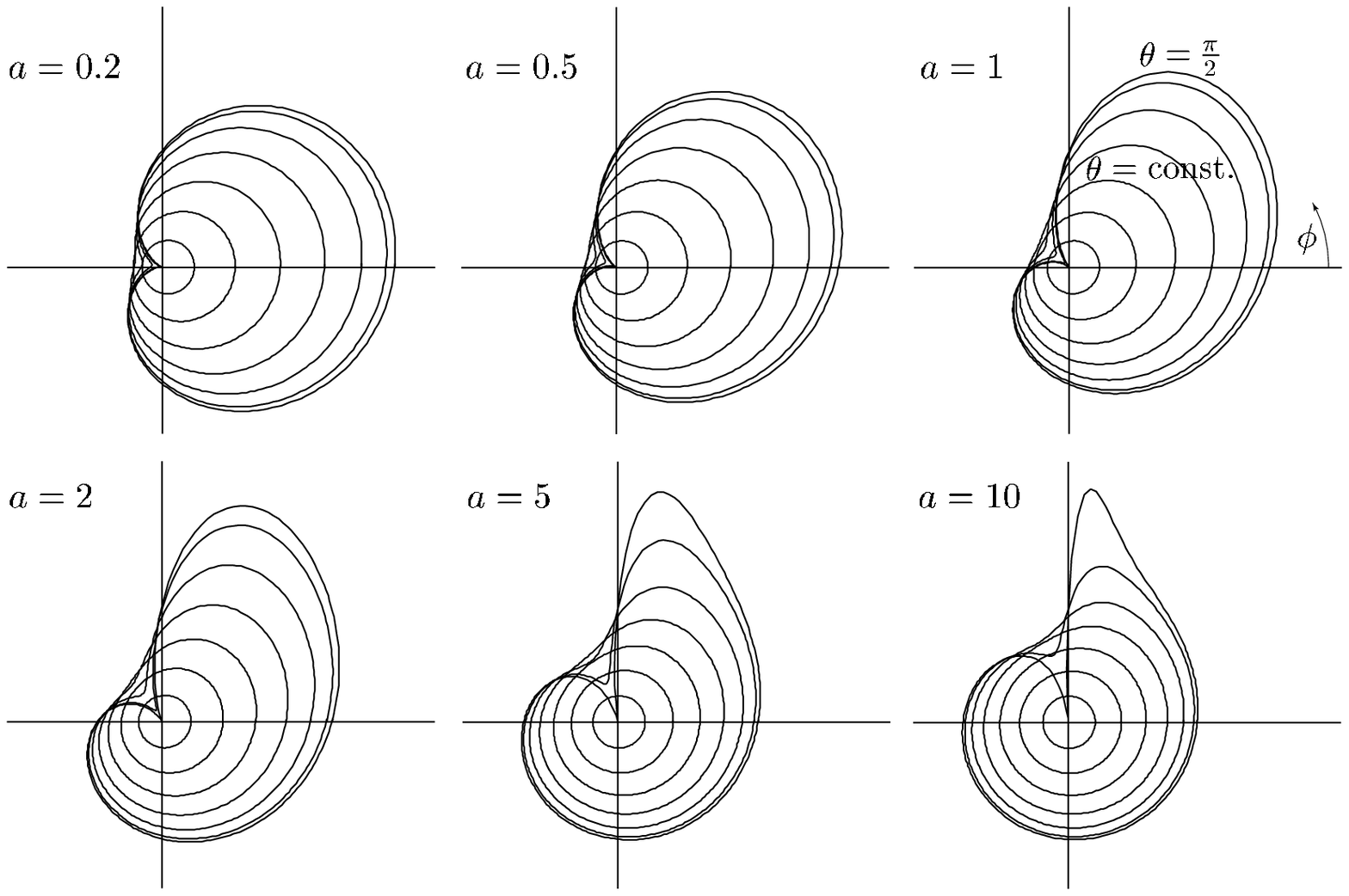}
   \end{center}
   \vspace{-5mm}
   \caption{\label{fig6}%
   Radiation patterns ${n^2(\theta,\phi)}$ of the photon field emitted by the rockets which move in four dimensions along circles of different radii~$a$ with the same proper angular velocity~$\omega$ (here ${\omega=1}$ and ${u=0}$). For each $a$ the curves plotted indicate the value of ${\,n^2(\theta=\hbox{const.},\phi)\,}$ in a polar graph, with the outer curve corresponding to the equatorial section ${\theta=\frac{\pi}{2}}$ (cf. figure~\ref{fig5} for the case ${a=1}$).  The asymmetry of the patterns grows for large~$a$ and thus velocities~$v$.}
\end{figure}

As an illustration, assuming ${D=4}$, the angular dependence of the emitted photon field on $\theta$ and $\phi$ (for a given value of the time $u$) is plotted as a spherical polar diagram in figure~\ref{fig5}.

Let us recall that, according to (\ref{pureradrhoD}), the radiation density $\rho$ decreases as ${\rho=r^{2-D}n^2(\theta_j,\phi)}$, where $r$ is the distance from the rocket, namely an affine parameter along null geodesics on ${u=\hbox{const.}}$ generated by ${\boldk=\partial_r}$.

In figure~\ref{fig6} we plot the radiation patterns (\ref{circpureradoptimal}) emitted at ${u=0}$ by the rockets which move along circles of different radii~$a$, keeping their proper angular velocity~$\omega$ fixed (${\omega=1}$). The curves shown correspond to ${\theta=\hbox{const.}}$, i.e., for a given $a$ and ${\theta}$ the value of ${n^2(\theta,\phi)}$ is plotted radially with $\phi$ being a standard polar angle. For larger $a$ --- and thus larger circular velocity $v$ given by (\ref{circvelocity}) --- the radiation patters become more distorted.

\begin{figure}[hb]
   \vspace{-1mm}
   \begin{center}
   \includegraphics[scale=0.78]{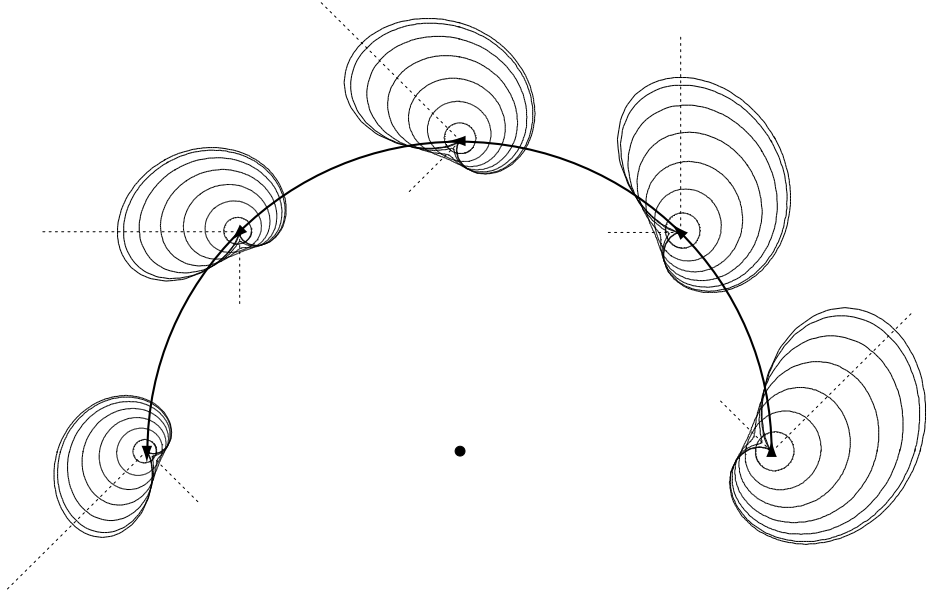}%
   \end{center}
   \vspace{-3mm}
   \caption{\label{fig7}%
   Radiation pattern ${n^2(\theta,\phi)}$ of the photon field emitted by the rocket at several places during its circular motion around the centre (${a\omega=1}$, ${D=4}$). The dashed lines indicate directions in which the radiation is maximal and minimal.}
\end{figure}

Due to the presence of the argument ${\,\phi-\omega\,u\,}$ in (\ref{circf}) and subsequent expressions, it is clear that the \emph{radiation pattern rotates} along the circular trajectory (\ref{circtraj}) with the proper angular velocity $\omega$. This is shown in figure~\ref{fig7} for the choice ${a\omega=1}$ and ${D=4}$.  Moreover, the radiation of photons \emph{decreases exponentially} with the factor ${\exp( -3\,a\omega^2\, u)}$, see (\ref{circpureradoptimal}), (\ref{circmoptimal}). In the proper scaling, the smallest pattern in figure~\ref{fig7} (on the left) should, in fact, be ${e^{3\pi}}$-times smaller than the largest one (on the right). Physically, since the mass of the rocket decreases exponentially, an exponentially decreasing ``reactive force'' from the emission of photons is sufficient to keep it on a circular motion with constant speed~$v$.

\begin{figure}[ht]
   \vspace{-2mm}
   \begin{center}
   \includegraphics[scale=0.75]{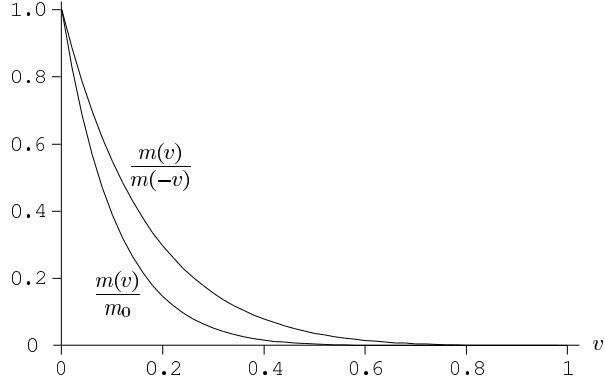}
   \end{center}
   \vspace{-5mm}
   \caption{\label{fig8}%
   Plot of the mass functions (\ref{massdescreasestraightvelocnew}) and (\ref{circmoptimalve}), assuming ${D=4}$.  The former is greater than the latter for any speed ${v\in(0,1)}$.}
   \end{figure}

We can also evaluate the mass function (\ref{circmoptimal}) after the ``U-turn'' depicted in figure~\ref{fig7}, using (\ref{circvelocity}) and the final time ${\omega\,u =\pi}$, as
 \begin{equation}
 \frac{m(v)}{m_0} =  \exp\left( -\frac{(D-1)\,\pi\,v}{\sqrt{1-v^2}}\right).
 \label{circmoptimalve}
 \end{equation}
Let us now compare this expression, which gives the mass loss after the \emph{circular manoeuvre} during which the speed $v$ of the photon rocket is kept constant while the direction of its flight is reverted, with the analogous relation corresponding to a constant deceleration ${\alpha}$, followed by the same acceleration, along a \emph{straight line}. Specifically, we employ the relation (\ref{massdescreasestraightveloc}) which gives the mass ${m(v)}$ of the rocket after its uniform acceleration along a straight line from ${v=0}$ to the speed $v$. However, this has to be combined with the initial straight flight during which the photon rocket decelerates from the velocity ${-v}$ to zero. By extending relations (\ref{motionstraight}), (\ref{massdescreasestraight}) and (\ref{massdescreasestraightveloc}) to negative values of $u$ and $v$ we obtain
 \begin{equation}
 \frac{m(v)}{m(-v)} = \left(\frac{1-v}{1+v}\right)^{D-1}\,.
 \label{massdescreasestraightvelocnew}
 \end{equation}
Since
 \begin{equation}
  -\frac{\pi\,v}{\sqrt{1-v^2}} \ < \ \log\left(\frac{1-v}{1+v}\right)
 \label{logcomp}
 \end{equation}
for any ${v\in(0,1)}$, the mass function (\ref{massdescreasestraightvelocnew}) is greater than (\ref{circmoptimalve}) --- see also the plot of these functions in figure~\ref{fig8}. In particular, for a small speed $v$,
 \begin{equation}
 \frac{m(v)}{m(-v)} \approx 1-2(D-1)\,v\,, \quad\hbox{while}\quad \frac{m(v)}{m_0}\approx 1-\pi(D-1)\,v\,.
 \label{massapprox}
 \end{equation}
We can thus conclude that in \emph{any} dimension $D$ the deceleration from the velocity~$v$ to zero and subsequent ``backward'' acceleration to the same velocity $v$ in exactly opposite direction along a straight line is more efficient than the circular U-turn of the rocket because the total mass of the photons emitted (the ``energy consumption'') is smaller.

\section{On motion of test particles and absence of ``gravitational aberration''}
\label{abersection}

In this final section we will briefly comment on geodesics which describe motion of free test particles in the exact spacetimes representing the gravitational field of an arbitrarily moving photon rocket.

To investigate such geodesics we have to evaluate the Christoffel symbols $\Gamma^\alpha_{\beta\gamma}$ for the metric (\ref{KerrSchild}). With respect to the Minkowski coordinates $Z^\alpha$, the metric reads
 \begin{equation}
 g_{\alpha\beta}=\eta_{\alpha\beta} + 2V\,  k_\alpha k_\beta ,\quad\ \hbox{so that}\quad\  g^{\alpha\beta}=\eta^{\alpha\beta} - 2V\,  k^\alpha k^\beta\,,
 \label{KerrSchildgeod}
 \end{equation}
where ${V(r,u)}$ is given by (\ref{functionV}). Moreover, using (\ref{transformation}) and (\ref{kunormaliz})--(\ref{kknormaliz}), for the functions ${u(Z^\alpha)}$ and ${r(Z^\alpha)}$ we obtain the relations
\begin{eqnarray}
 u_{,\mu} &=& -k_\mu\,,  \nonumber\\
 r_{,\mu} &=& (1+rk_\sigma\ddot z^\sigma)\,k_\mu - \dot z_\mu\,, \label{auxirel}\\
 (rk_\alpha)_{,\mu} &=& \eta_{\alpha\mu}+\dot z_\alpha \,k_\mu\,,\nonumber
\end{eqnarray}
where ${\dot z_\alpha \equiv \eta_{\alpha\beta}\, \dot z^\beta }$. Straightforward but somewhat lengthy calculation then yields
\begin{eqnarray}
&&\hspace{-12mm}
\Gamma^\alpha_{\beta\gamma} = \frac{2V}{r}\,k^\alpha\,\eta_{\beta\gamma} + V_{,r}\,\dot z^\alpha k_\beta k_\gamma+\big(2VV_{,r}-V_{,u}\big)\,k^\alpha k_\beta k_\gamma \nonumber\\
 &&\hspace{-2mm} +\Big(\frac{2V}{r}-V_{,r}\Big)\big(k^\alpha\dot z_\beta k_\gamma +k^\alpha\dot z_\gamma k_\beta-(1+rk_\sigma\ddot z^\sigma)\,k^\alpha k_\beta k_\gamma\big)\,. \label{Christof}
\end{eqnarray}
Considering ${\,k^0=-k_0=p^{-1}\,}$, ${\,k^i=k_i=p^{-1}n^i\,}$ where $p$ is given by (\ref{functionP}) and ${\boldn=n^i\partial_i}$ is the spatial unit vector (see section~\ref{trajMink}), all components of ${\Gamma^\alpha_{\beta\gamma}}$ can be written explicitly.

It is also convenient to introduce a spatial velocity vector ${\boldv=v^i\partial_i}$ of the rocket with respect to the Minkowski background frame by ${v^i\equiv \dot z^i(u)/\dot z^0(u)}$. Consequently,
 \begin{equation}
 \dot z^0(u)=\gamma\,,\qquad
 \dot z^i(u)=\gamma\,v^i\,,\qquad \hbox{where}\quad
 \gamma=\frac{1}{\sqrt{1-\boldv\cdot\boldv}}
 \label{splitting}
 \end{equation}
is the standard Lorentz factor corresponding to the velocity $\boldv$ of the rocket at ${\cal P}$ (i.e., at its proper time $u$), see figure~\ref{fig1}. The function $p$ then simplifies to
 \begin{equation}
 p=\gamma \,(1-\boldn\cdot\boldv) \,.
 \label{pfunc}
 \end{equation}
Let us recall that $\boldn$ is the unit vector connecting the (spatial position of) events ${\cal P}$ and ${\cal Q}$. In fact, the \emph{spatial distance} $R$ between ${{\cal P}\equiv\{z^0(u), \ldots, z^{D-1}(u)\}}$ and ${{\cal Q}\equiv\{Z^0, \ldots, Z^{D-1}\}}$ is ${R=\sqrt{(Z^1-z^1(u))^2+\cdots+(Z^{D-1}-z^{D-1}(u))^2}}$, which using (\ref{transformation}), (\ref{kknormaliz}) becomes ${\,R=rk^0}$. It follows that
 \begin{equation}
 r=p\,R\,.
 \label{distance}
 \end{equation}
From (\ref{Christof}), (\ref{functionV}) we thus obtain
\begin{eqnarray}
&&\hspace{-22mm}
\Gamma^i_{00} = -\left(\frac{2m}{R^{D-2}p^{D-1}}+\frac{2\Lambda\,R}{(D-2)(D-1)}\right)\!\!\left(1+\frac{(D-3)\,m}{R^{D-3}p^{D-1}}-\frac{2\Lambda\,R^2}{(D-2)(D-1)}\right)n^i  \nonumber\\
 &&\hspace{-13mm} +\left(\frac{(3-D)\,m}{R^{D-2}p^{D-1}}+\frac{2\Lambda\,R}{(D-2)(D-1)}\right)\frac{\gamma\, v^i}{p}\nonumber\\
 &&\hspace{-13mm} +\left(\frac{(D-1)(2\gamma p-1)\,m}{R^{D-2}p^{D+1}}-\frac{(D-1)\,m\,k_\sigma\ddot z^\sigma + \dot m}{R^{D-3}p^D}\right)n^i. \label{Christofi00}
\end{eqnarray}
Due to the geodesic equation, the coefficient ${-\Gamma^i_{00}}$ gives the ``spatial acceleration'' (in Newtonian terminology) of a test particle which is at rest at a given point~${\cal Q}$. Obviously, for large distances $R$ of the test particle from the photon rocket, the dominant contribution to its acceleration arises from the cosmological constant $\Lambda$ which represents global isotropic expansion of the spacetime when ${\Lambda>0}$. On the other hand, for small ${R}$ the contribution of the $\Lambda$-terms is negligible.

Setting ${\Lambda=0}$ in (\ref{Christofi00}) and employing relations (\ref{splitting}), (\ref{pfunc}) we obtain
\begin{eqnarray}
&&\hspace{-22mm}
\Gamma^i_{00} = \frac{m}{R^{D-2}\gamma^{D-1}\,(1-\boldn\cdot\boldv)^{D+1}}\bigg[\Big( (D-3)-2(D-3)\,\boldn\cdot\boldv-2\,(\boldn\cdot\boldv)^2  \nonumber\\
 &&\hspace{-13mm} +(D-1)\,\boldv\cdot\boldv \Big)n^i - (D-3)(1-\boldn\cdot\boldv)\,v^i - \frac{2(D-3)\,m}{R^{D-3}\gamma^{D-1}(1-\boldn\cdot\boldv)^{D-3}}\,n^i  \bigg] \nonumber\\
 &&\hspace{-13mm} -\frac{(D-1)\,m\,k_\sigma\ddot z^\sigma + \dot m}{R^{D-3}\gamma^D (1-\boldn\cdot\boldv)^D }\,n^i. \label{Christofi00apr}
\end{eqnarray}
The last term, proportional to the acceleration ${\ddot z^\sigma }$ and mass decrease ${\dot m}$ of the rocket, represents the radiative part of the gravitational field which behaves as ${\propto R^{3-D}}$. Close to the photon rocket this can be neglected with respect to the first ``Newtonian force''-term ${\propto R^{2-D}}$. Notice that such Newtonian-like gravitational acceleration is oriented along
 \begin{equation}
 (D-3)\big[ (n^i-v^i)-(2n^i-v^i)\,\boldn\cdot\boldv + \ldots \big]\,,
 \label{direction}
 \end{equation}
which is the spatial direction toward the ``instantaneous'' position of the rocket, extrapolated from its ``retarded'' position given simply by $n^i$. This demonstrates the \emph{absence of ``gravitational aberration''} in such systems which are explicit exact solutions of Einstein's equations in any dimension $D$. For a thorough discussion of this issue in ${D=4}$ see the Carlip work~\cite{Car00}.

\section{Conclusions}
\label{RTfinalsection}

We analysed in detail some properties of the class of exact spacetimes which represent Kinnersley's photon rockets moving arbitrarily in any dimension~$D$. These solutions contain pure-radiation photon field and admit a cosmological constant~$\Lambda$. They are of algebraic type~D and can be written either in the Kerr--Schild form (\ref{KerrSchild}), (\ref{functionV}) or more explicitly in the Robinson--Trautman forms (\ref{geo_metric finF}), (\ref{functionP2cartF}) and (\ref{RTMinkowskiDspherF}), (\ref{functionPDspherF}).

In section~\ref{trajMink} we first systematically reviewed general Newman--Unti [(\ref{RTMinkowskiD}), (\ref{functionP})], spherical [(\ref{RTMinkowski2spher}), (\ref{functionP2spher}), or (\ref{RTMinkowski2stereo}), (\ref{functionP2stereo}), or (\ref{RTMinkowskiDspher}), (\ref{functionPDspher})] and Cartesian-like [(\ref{RTMinkowski2cart})--(\ref{functionP2cart})] background coordinates suitable for the description of an arbitrarily moving test source in flat Minkowski space of dimension~$D$.

The complete Kinnersley solution in various metric forms was presented and discussed in subsequent section~\ref{propsection}. A special case of photon rockets accelerating in a single spatial direction is contained in section~\ref{singlesection}.

We discussed important particular trajectories of the rockets, namely a straight flight (in subsection~\ref{singlesectionstraight}) and a circular motion (in subsection~\ref{circular}), including the corresponding radiation patterns of the photon field and the mass loss formulae. For example, we showed that the straight deceleration followed by the backward acceleration is a more efficient manoeuvre then the circular U-turn since the total mass of the emitted photons (and thus the energy consumption) is smaller in the former case.

In the final section~\ref{abersection} we derived the Christoffel symbols for general Kinnersley spacetimes which are crucial for discussion of geodesics. In particular, we demonstrated that, in any dimension $D$, there is no ``gravitational aberration'' effect.

To conclude, let us remark that in our contribution (as well as in previous works \cite{Bon94,Dam95,Bon96,DaiMorGle96,CorMic96,vonGonKram98,Cor00,Car00,Pod08}) the Kinnersley solution has been used as an exact model of an accelerating photon rocket. However, in ${D=4}$, it may also be considered as a simple relativistic model of non-geodetically moving astronomical object in the Solar System (such as a small particle or an asteroid) which accelerates due to a specific anisotropic thermal absorption/emission of photons, i.e., to simulate the Poynting--Robertson effect \cite{BurLamSot79} or the Yarkovsky effect \cite{BotVokRubNes06}.

\section*{Acknowledgements}

This work was supported by the grant GA\v{C}R~202/08/0187, the Czech Centre for Theoretical Astrophysics LC06014 and by the project MSM0021620860.

I would also like to thank the Icelandic volcano Eyjafjallaj\"okull to give me more time, during the leave of absence at Loughborough University, to elaborate this paper.

\end{document}